\documentclass[fleqn,usenatbib]{mnras}

\usepackage{newtxtext,newtxmath}

\usepackage[T1]{fontenc}
\usepackage{bold-extra}
\usepackage{ae,aecompl}
\usepackage{soul}
\DeclareRobustCommand{\VAN}[3]{#2}
\let\VANthebibliography\thebibliography
\def\thebibliography{\DeclareRobustCommand{\VAN}[3]{##3}\VANthebibliography}

\usepackage{pdflscape} 

\usepackage{graphicx}	
\usepackage{amsmath}	

\usepackage[normalem]{ulem}
\usepackage[]{xcolor}
\usepackage{comment}
\usepackage[font=small,skip=0pt]{caption}

\newcommand{\nbodysix}{{\sc nbody6++gpu}}

\usepackage{csquotes}
\usepackage{newtxtext,newtxmath}

\title[GC disruption in dwarf galaxies]{The influence of globular cluster evolution on the specific frequency in dwarf galaxies}

\author[E. Moreno-Hilario et al.]{
Elizabeth Moreno-Hilario$^{1}$\thanks{E-mail:\href{mailto:emoreno@astro.unam.mx}{ emoreno@astro.unam.mx}}\href{https://orcid.org/0000-0002-6906-2379}{\includegraphics[scale=0.5]{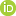}},
Luis A. Martinez-Medina$^{1}$\thanks{E-mail:\href{mailto:lamartinez@astro.unam.mx}{ lamartinez@astro.unam.mx}}\href{https://orcid.org/0000-0002-5749-8255}{\includegraphics[scale=0.5]{orcid.png}},
Hui Li$^{2,3}$\thanks{E-mail:\href{hliastro@tsinghua.edu.cn}{ hliastro@tsinghua.edu.cn}}\href{https://orcid.org/0000-0002-1253-2763}{\includegraphics[scale=0.5]{orcid.png}},
Stefano O. Souza$^{4}$\href{https://orcid.org/0000-0001-8052-969X}{\includegraphics[scale=0.5]{orcid.png}}, and 
\newauthor
Angeles Pérez-Villegas$^{5}$\href{https://orcid.org/0000-0002-5974-3998}{\includegraphics[scale=0.5]{orcid.png}} 
\\
$^{1}$Instituto de Astronomía, Universidad Nacional Autónoma de México, A. P. 70-264, C.P. 04510, CDMX, México\\
$^{2}$Department of Astronomy, Tsinghua University, Beijing 100084, China\\
$^{3}$Department of Astronomy, Columbia University, New York, NY 10027, USA\\
$^{4}$Universidade de São Paulo, IAG, Rua do Matão 1226, Cidade Universitária, São Paulo 05508-900, Brazil\\
$^{5}$Instituto de Astronomía, Universidad Nacional Autónoma de México, A. P. 106, C.P. 22800, Ensenada, B. C., México\\
}

\date{Accepted XXX. Received YYY; in original form ZZZ}

\pubyear{2021}

\begin{document}
\label{firstpage}
\pagerange{\pageref{firstpage}--\pageref{lastpage}}
\maketitle

\begin{abstract}

Dwarf galaxies are known to exhibit an unusual richness in numbers of globular clusters (GCs), property quantified by the specific frequency ($S_N$), which is high for dwarf and giant elliptical galaxies, but with a minimum for intermediate-mass galaxies. In this work we study the role that GC evolution has in setting this trend, for which we use {\it N}-body simulations to evolve GCs in dwarf galaxies and quantify their disruption efficiency. We selected five individual dwarf galaxies from a high-resolution cosmological simulation, which includes GC formation and follow-up of their paths inside the host galaxy. Then, the tidal history of each GC is coupled to {\nbodysix} to produce {\it N}-body models that account for both, the interaction of GCs with their galactic environment and their internal dynamics. This results in a GC mass loss parameterization to estimate dissolution times and mass loss rates after a Hubble time. GC evolution is sensitive to the particular orbital histories within each galaxy, but the overall result is that the amount of mass that GC systems lose scales with the mass (and density) of the host galaxy, i.e., the GC mass loss efficiency is lowest in low-mass dwarfs. 
After a 12 Gyr evolution all simulated GC systems retain an important fraction of their initial mass (up to 25\%), in agreement with the high GC to field star ratios observed in some dwarfs, and supports the scenario in which GC disruption mechanisms play an important role in shaping the GC specific frequency in dwarf galaxies.

\end{abstract}

\begin{keywords}
galaxies: star clusters: general --- globular clusters: general --- galaxies: dwarf --- galaxies: evolution
\end{keywords}



\section{Introduction}
\label{section:Introduction}
Globular Clusters (GCs) are gravitationally bound stellar systems, formed in high-redshift environments around the epoch of reionization \citep{Forbes2018,Mowla2022,Vanzella2023}, with a fraction of them dense enough to endure the merger history of their host galaxy. They can be found in all galaxy morphological types, covering a wide range of galactic stellar masses, from giant elliptical (M$_*$ $ \approx 10^{12}$ M$_\odot$) to dwarf (M$_* \leq 10^9$ M$_\odot$) galaxies \citep[e.g.][]{Harris1996AJ,Peng2006,Harris2016,Harris2017}. 

Being born within sites of active star formation, GCs experienced an early mass loss due to expulsion of remnant gas through feedback and tidal shocks by dense molecular structures in the star-forming environment \citep{1999ApJ...513..626G, 2007MNRAS.376..809G, 2008ApJ...689..919P, 2019ARA&A..57..227K,2019MNRAS.487..364L}. Mass loss continued for the clusters that endured such hostile conditions, dominated by evaporation and tidal disturbances, where long-term evolution depends on the environment and perturbations that the cluster experiences along its path \citep[e.g.][]{Spitzer1958,1972ApJ...176L..51O, 1988ApJ...335..720A, 2009ApJ...706...67R,MartinezMedina2022,2022MNRAS.514..265L}.

Global properties of GC systems such as population size ($N_{\rm GC}$), metallicity, age distribution, spatial arrangement, mass function, specific frequency, dynamical characteristics, and radial metallicity gradient have been extensively studied, as they keep record of the formation and evolution history of their host galaxies \citep{1991ApJ...379..157B, 2001AJ....121.2974L, 2001MNRAS.322..257F, 2004AJ....127.3431S, 2006ARA&A..44..193B, 2020rfma.book..245B, 2021IAUS..359..381B}.

Such properties provide valuable insights into galaxy assembly, chemical enrichment, and the mechanisms influencing GC formation and survival over cosmic time. These properties also scale with the mass of the host galaxies, and of particular interest is the specific frequency ($S_N$), defined as the number of GCs per unit of galaxy luminosity \citep[$M_V$;][]{1981AJ.....86.1627H}; it was introduced as a diagnosis to measure the richness of GCs systems in galaxies. This quantity is particularly high in dwarf galaxies; it increases as the galaxy luminosity decreases and can reach values as large as $S_N>100$ \citep{2010MNRAS.406.1967G}, while in spiral galaxies this value typically is $S_N =1$ \citep{Harris2013, 2000ApJ...533..869C}. Such large discrepancies among $S_N$ values are often called the specific frequency problem.

Explanations of the observed trends in $S_N$ have been directly related to the efficiency of galaxies to form GCs \citep[e.g.][]{Peng2008,Harris2013,Larsen2012,Larsen2014,Choksi2018}. The general picture is that GCs were born first, during events of high star formation (SF), after which the formation of field stars proceeds. It turns out that if the SF that followed after GC formation was suppressed or diminished, then the $S_N$ values would have become large. 
A series of mechanisms have been proposed to be capable of diminishing field SF. In the regime of high galactic masses, active galactic nucleus (AGN) feedback and virial shock heating due to infalling gas can reduce the SF efficiency \citep{1981AJ.....86.1627H}; while SF in dwarf galaxies is more prone to be suppressed by supernova feedback \citep[][]{Dekel1986,Dekel2003}. On the other hand, these quenching mechanisms are expected to be less efficient for intermediate-mass galaxies, which would reduce the GC to field star ratios and hence the specific frequency. Moreover, in a recent work, \citet{Andersson2023} use hidrodynamical simulations of a dwarf galaxy to show that star cluster formation efficiency is sensitive to different feedback mechanisms. In particular, the authors find that the timing of feedback initiation shapes the initial cluster mass function, preferentially affecting the number of formed massive clusters. 

Another mechanism proposed as capable of shaping the specific frequency is GC disruption. This idea was first explored through numerical simulations by \citet{Murali1997}, where they evolve GC populations in elliptical galaxies using the Fokker-Plank approximation. Their main result is that GC depletion scales with galaxy luminosity, being larger in denser, less luminous ellipticals, reproducing the observed $S_N$ values in the explored galaxy mass range. Following a similar approach, \citet{2014A&A...565L...6M} studied the U-shaped specific frequency as the outcome of tidal erosion of GCs. To quantify this effect they analysed a set of simulations presented previously in \citet{2014MNRAS.441..150B}, and computed survival rates of GCs in spherical galaxy models. With their methodology, they also found that the erosion of GCs plays an important role in setting the U-shape relation of the specific frequency as a function of galaxy luminosity.

Moreover, the environment in which galaxies evolve also has an effect on the GC formation efficiency. \citet{Mistani2016} used the Illustris simulation to study the properties of dwarf galaxies in different environments. They identify two sets of dwarfs; dwarf elliptical galaxies (dEs), typical of galaxy clusters, and gas-rich dwarf irregulars (dIrrs), generally found in the field. The authors found that dwarfs in clusters present important starburst events at early times, triggered mainly by the infall of the galaxy into the cluster. This gives a natural explanation for the observational evidence of dEs in clusters that show specific frequency values systematically larger than dIrrs in the same mass range \citep{Miller2007,Jordan2007,Peng2008}.

However, a second specific frequency problem emerges when the number of GCs is compared to their corresponding stellar population in the host galaxy. Deriving the abundance patterns of GCs in the Fornax, Wolf-Lundmark-Melotte (WLM), and IKN dwarf galaxies, \citet{Larsen2012,Larsen2014} estimated the ratios of metal-poor GCs to field stars, and found that large fractions of low-metallicity stars ([Fe/H]<-2.0) in these galaxies belong to GCs; such fractions are noticeable high when compared with the Milky Way (MW) halo stars. 
These ratios constrain the amount of mass lost by metal-poor GCs in dwarf galaxies as a consequence of their early evolution and subsequent dynamical disruption, implying that the initial mass of these clusters was just a factor of few larger than the current value.

In this work, we study the GC to field star ratios in dwarf galaxies from a dynamical point of view. By modelling $N$-body star clusters evolving inside dwarfs, we compute the mass lost by GC systems as a function of their host's mass. This allows us to look into the contribution that GC disruption processes have on the specific frequency trends at low metallicities, and understand why it appears that GCs in dwarfs have lost only small fractions of their initial mass. To model the disruption of an $N$-body star cluster inside an evolving galaxy we adopt the methodology used in \citet{Renaud2011} and \citet{Rieder2013}, which, as we mention in the following sections, consists on describing the cluster on the one hand and its environment on the other, with the purpose of coupling these two aspects.

This paper is organised as follows. In Section \ref{section:GCs_Simulations}, we present a brief description of the cosmological simulation from which the dwarf galaxies were extracted, and the $N$-body code that couples these galaxies with the GCs. In Section \ref{section:MassEvolutionGCs}, we compute and characterise the mass loss of GCs, as well as their mass loss rate as a function of the host galaxy mass. Then, we evolve GC systems in each of our dwarf galaxies and compare the mass loss of these systems. In Section \ref{section:Discussion}, we discuss this mass evolution in the context of the second specific frequency problem. Finally, in Section \ref{section:Conclusions} we present our conclusions.

\section{Globular Cluster Simulations}
\label{section:GCs_Simulations}

\subsection{Cosmological Simulations}
\label{subsection:CosmoSim}

A detailed description of the simulations setup and implementation of the cluster formation algorithm is presented in \citet{li_etal17} and \citet{2018ApJ...861..107L}. Here, we briefly describe some key features of the cosmological simulations. Different from other cosmological simulations, \citet{li_etal17} developed a new prescription by considering star clusters as a unit of star formation in the mesh-based Adaptive Refinement Tree (ART) code. In this model, after a cluster particle is seeded at a density peak, it grows in mass continuously via gas accretion from its neighbouring cells. To mimic the behaviour of real star clusters, the mass growth is resolved with high time resolution and is terminated by its own energy and momentum feedback; thus, the final cluster mass is set self-consistently. After the cluster particles emerge from their natal clouds, they orbit around the host galaxies and experience mass loss caused by stellar evolution and tidal disruption which is described in the next subsection. This series of works reproduces various properties of young massive clusters, such as the shape of the initial cluster mass function and cluster formation efficiency.

From this cosmological simulation we selected five dwarf galaxies, labelled as galaxy A, B, C, D, and E, going from the least to the most massive. These galaxies exhibit a range of dynamical masses from $10^9$ to $10^{11}$ M$_\odot$, and stellar masses from $5 \times 10^{6}$ to $5 \times 10^8$ M$_\odot$. Each galaxy mass distribution plays a crucial role in shaping its gravitational potential, and this potential influences the motion and behaviour of objects within it, e.g., the dynamics of GCs.

\subsection{Tidal Tensors}
\label{subsection:TidalTensors}

The interaction between GCs and their host galaxy can be expressed in the form of tidal tensors. The tensors are obtained from the high-resolution cosmological simulations described in the previous section, which is the same as used in \citep{Li2019}.

The tidal tensor, defined as the second derivative of the gravitational potential of the cluster's host galaxy $\Phi_G$
\begin{equation}
\label{eq:T}
T_{ij} = -\frac{\partial^2\phi_G}{\partial x_i\partial x_j},
\end{equation}
contains the tidal interactions that the cluster experiences due to its environment. Within the simulated galaxies, GC particles are identified, and the corresponding tidal tensor is calculated along their orbit. 

Since the galactic environment evolves during the simulation, the gravitational potential of the cluster's host galaxy is time-dependent, with no analytical expression for $\Phi_G$ in eq. \ref{eq:T}; hence, the tidal tensor must be computed numerically. It is estimated using a second-order finite difference scheme, over the scale of 24 pc \citep{Li2019}. This scale is comparable to the estimates of tidal radii of the Galactic GCs, which are of the order $20 - 50$ pc. Thus, this scheme is appropriate for capturing the tidal field around the model clusters.

Once the tidal tensors are computed for randomly selected clusters in each galaxy, we calculate their three eigenvalues, $\lambda_1 \geq \lambda_2 \geq \lambda_3$, and the corresponding eigenvectors. The eigenvalues give the magnitude of the tidal field, whereas the eigenvectors give the direction along which the system is stretched or compressed.

\begin{figure}
	\includegraphics[width=\columnwidth, trim=20 10 10 0]{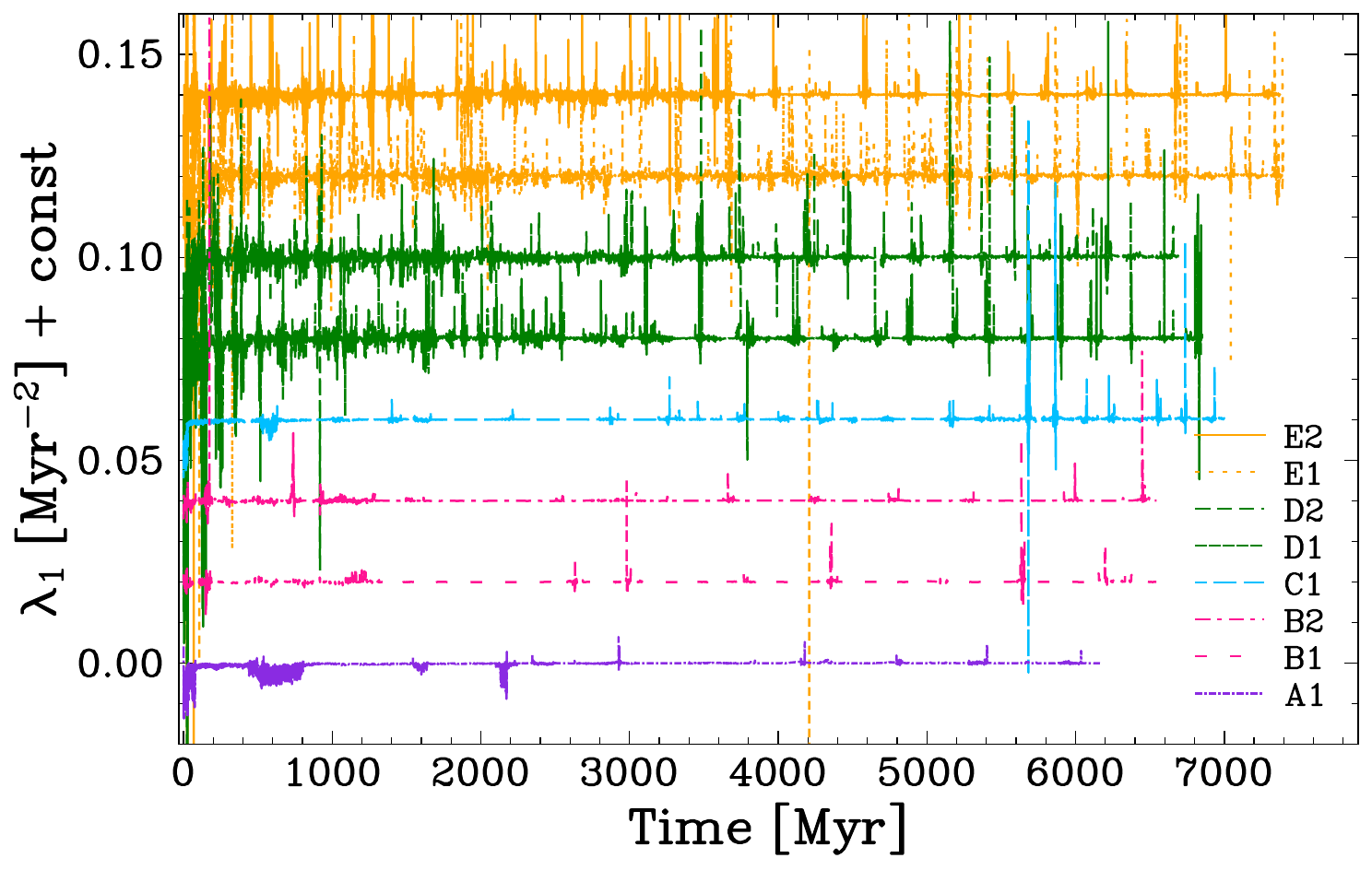}
	\centering
    \caption{Maximum eigenvalues of the tidal tensor as a function of time. The color groups correspond each to a different satellite galaxy in the cosmological simulation. To facilitate easy differentiation between the tensors, a Y-shift has been applied to each line.}
    \label{fig:TidalStrenght}
\end{figure}

The maximum eigenvalue, $\lambda_1$, provides insight into the strength of the tidal field. In our analysis, we consider a set of tensors labelled as A, B, C, D, and E, which correspond to the labels used for the galaxies they are extracted from. In addition, to distinguish among the tensors within each galaxy we enumerate them accordingly. Figure~\ref{fig:TidalStrenght} shows the $\lambda_1$ values for the tensors we are working with, and the colours are used to identify the different galaxies where the GCs were formed in the cosmological simulation. It is notable that the tensors associated with more massive galaxies exhibit more frequent higher amplitudes, implying a more violent environment. This is particularly evident in the tensors labelled as ``D1'', ``D2'', ``E1'', and ``E2''. These strong interactions extend beyond the early stages and persist throughout several Gyr, indicating an ongoing turbulent environment. Consequently, GCs within these more massive galaxies are subject to significant mass loss and experience violent tidal forces throughout their evolutionary history. Moreover, most tensors exhibit strong tidal forces during the cluster formation phase, probably influenced by the high-density environment in which they originated.

In general, the tensors are quite different from each other, even if they come from the same satellite galaxy, some exhibit peaks and oscillations during most of the simulation, probably due to cluster encounters with giant molecular clouds (GMCs) or disc crossings; while others are relatively quiet after the first 500 Myr. With such differences, we can notice that the strength of the tides depends on the galactic environment. This behaviour was studied by \cite{2008MNRAS.389L..28G}, where they carried out an extensive survey of $N$-body simulations showing that, for a cluster with a given number of particles (i.e. stars), the strength of the tidal field is the dominant factor that determines the lifetime of a cluster.

\subsection{{\it N}-body Code}
\label{subsection:Nbody6}

Once the tidal tensors are computed, we couple them with the $N$-body code {\nbodysix}\footnote{\url{https://github.com/nbodyx/Nbody6ppGPU.git}} \citep{nitadori_aarseth12} which has been specifically designed for star clusters and includes the treatment of galactic tides, taking tidal tensors as an input. This code allows us to simulate the cluster's stellar evolution, while also accounting for mass loss resulting from two-body interactions.
{\nbodysix} creates an $N$-body model of a cluster, star by star, taking the tidal tensors as an external force, calculating the tidal forces at the positions of the $N$ stars in the cluster, and adding them to the internal gravitational force due to the remaining $N-1$ stars. A detailed description of the features of {\nbodysix} are presented in \citet{Renaud2011}, \citet{2015MNRAS.448.3416R} and \citet{2015MNRAS.450.4070W}. Here, we briefly introduce some key features of the code.

{\nbodysix} uses the fourth-order Hermite integration method. To speed up force calculation, the Ahmad-Cohen (AC) neighbour scheme is used for integration \citep{1973JCoPh..12..389A}, where the basic idea is to employ a neighbour list for each particle. Then the integration is separated in regular force for large time steps (regular steps), and irregular force for small time steps (irregular steps). The regular force is the sum of the forces from particles outside the neighbour radius, and the irregular force accumulates only the neighbour forces. 

A key ingredient in modelling $N$-body star clusters are black holes (BHs), which result from stellar evolution. While their presence in GCs was debated until recently, \cite{2007Natur.445..183M} confirmed a BH in a GC near NGC 4472. This prompted questions about the role of undetected BHs in GC dynamics. In a recent study, \cite{2021NatAs...5..957G} explored Palomar 5 (Pal 5) formation and evolution, using $N$-body simulations with and without BHs. BH-inclusive simulations showed clusters expanding, resulting in less concentrated observable stars compared to dynamical mass. Models lacking BHs required specific initial conditions, indicating lower likelihood. This research underscores BHs' potential impact on GC dynamics.

To properly model BH numbers, the latest {\nbodysix} version incorporates a new BH retention treatment \citep[see][]{2020A&A...639A..41B}. It includes updated modelling of natal kicks on neutron stars and BHs, along with momentum-conserving natal kicks that dependent on the supernova material fallback.

\subsection{Initial Conditions for Globular Clusters}
\label{subsection:Simulations}

In this work, we present the results for a total of 24 simulations with different galactic environments which are given by the tidal tensor. To ensure a robust extrapolation of GC mass loss rates, especially for massive GCs, we chose to have three sets of 8 $N$-body simulations, with different initial number of particles ($N_0$); for the first set $N_0 = 32000$ (hereafter referred to as set 1), $N_0 = 64000$ for the second set (hereafter, set 2) and $N_0 = 96000$ for the third set (hereafter, set 3). Having at least three initial masses allows us to make a more reliable estimation. Our sets of simulations consist of the following properties:

\begin{itemize}
    \item The GCs start off with a total initial mass of $M_0 = 20,480$ M$_\odot$, $M_0 = 40,960$ M$_\odot$ and $M_0 = 61,440$ M$_\odot$ for GCs in set 1,  set 2 and set 3, respectively.
    \item Plummer model is used as the initial density profile of the clusters.
    \item Stellar initial mass function (IMF) of \citet{2001MNRAS.322..231K}, set in a range 0.08 – 100.0 M$_\odot$.
    \item Initial mass density within half-mass radius of $\rho_{\rm h,0} = 10^3$ M$_\odot/$pc$^3$.
    \item Half-mass radius of $r_{\rm h,0} = (1.34, 1.69, 1.94)$ pc for GCs in set 1, set 2 and set 3, respectively.
\end{itemize}

The exploration of other initial density values and tidal tensors is out of the scope of this work, mainly due to the computation time required to create a set of $N$-body models. However, in Section \ref{section:Discussion} we discuss the effects of adopting different initial cluster densities.

\section{Mass evolution of GC\lowercase{s}}
\label{section:MassEvolutionGCs}

\begin{figure}
	\includegraphics[width=\columnwidth, trim=20 10 10 0]{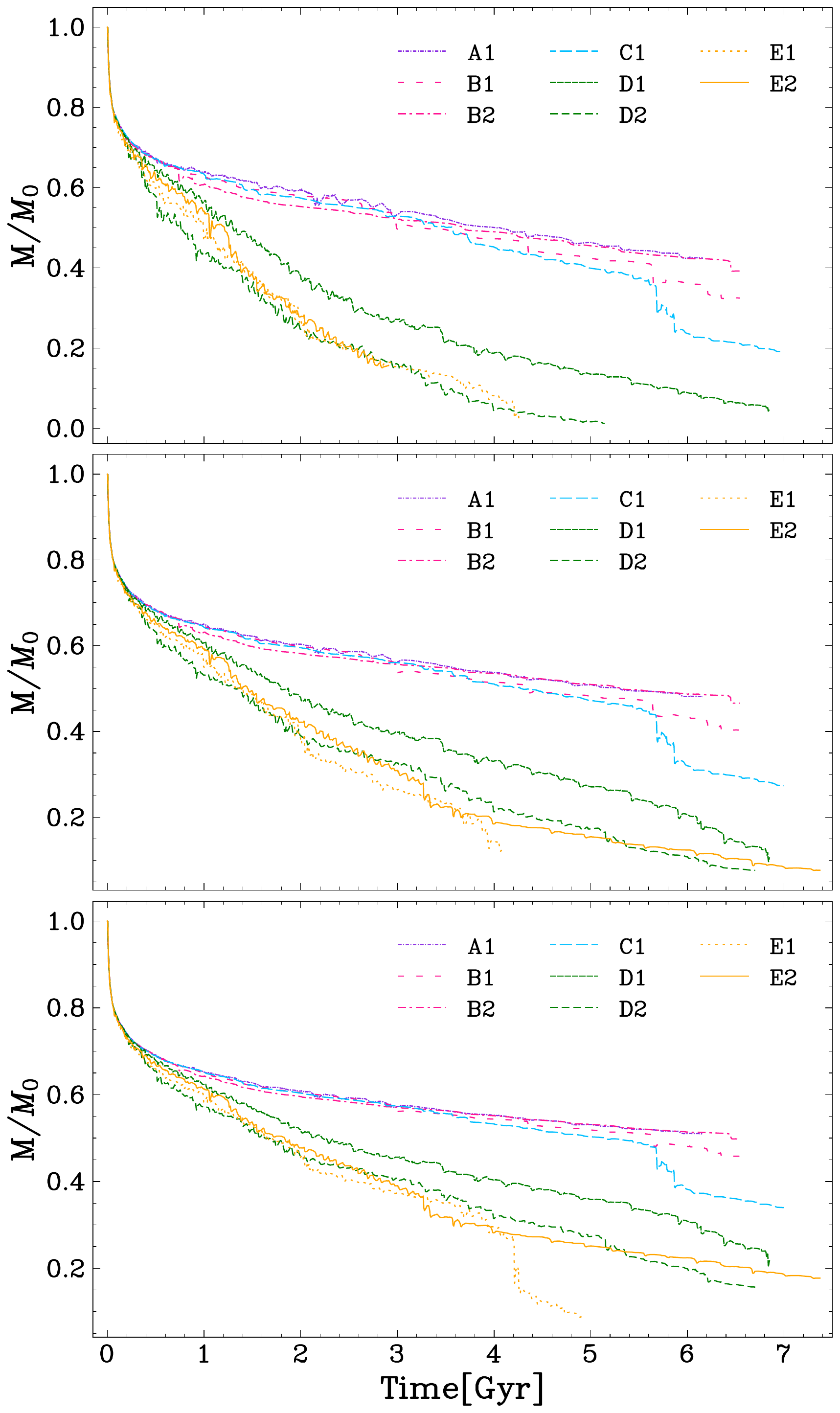}
	\centering
    \caption{Evolution of cluster masses over time in the three sets of simulations: set 1 (top panel), set 2 (middle panel) and set 3 (bottom panel). The colour code corresponds to that of the tensor used as the external tidal force of each GC simulation as shown in Figure~\ref{fig:TidalStrenght}.}
    \label{fig:MassEvolution_sets}
\end{figure}

In Figure~\ref{fig:MassEvolution_sets} we show the mass evolution of our simulated clusters over time. Initially, all clusters show a similar evolution in the first few Myr, characterised by a drop in mass driven by stellar evolution that, for the IMF used here, results in an average reduction of $\sim 30\% $ in the initial GC mass. This early mass loss process has been well studied and characterised in the literature \citep[e.g.,][]{2003MNRAS.340..227B, 2008ApJ...689..919P, 2010MNRAS.409..305L}. Beyond this initial phase, GCs exhibit diverse evolutionary paths; some clusters experience rapid disruption, while others show a stepped pattern of significant mass reduction before continuing to evolve more smoothly. Thus, as time progresses, the dominant mechanisms for mass loss shift to dynamical processes, which play a crucial role in shaping the cluster's evolution throughout its lifetime.

In general, the simulations show that GCs in set 3 have longer lifetimes compared to those in sets 1 and 2. Consequently, there is a greater likelihood of complete dissolution for low-mass GCs. Notably, the differences in survival rates among the sets are particularly prominent for the clusters evolved with tensors D1, D2, E1, and E2. As shown in Figure~\ref{fig:MassEvolution_sets}, in these clusters it is more evident that they live longer as the initial mass of the clusters increases.

Another thing to note is that the evolution of the cluster depends on the tidal tensor, that is, depends on the orbit and galactic environment, and we can see that the lifetimes of clusters can vary significantly. At the times when the tidal tensor presents higher variations, the cluster shows a sudden mass loss. Depending on the mass of the cluster, these high variations of the galactic environment can lead to complete disruption of clusters, while a more massive cluster can survive this interaction with the host galaxy.

\subsection{Mass Loss Models of GCs}
\label{subsection:MassLossModels}

The mass of star clusters decreases during their lifetime until they are finally completely dissolved. As we already observe in the results depicted in Figure~\ref{fig:MassEvolution_sets}, stellar evolution dominates the mass loss only in the early phase of the cluster lifetime, and it is already taken into account in our simulations with \nbodysix{}.

On the other hand, following the works by \citet{2010MNRAS.409..305L}, \citet{2001MNRAS.325.1323B}, and \citet{2021MNRAS.503.3000M}, the time-dependent mass loss by dynamical effects can be described by
\begin{equation}
\label{eq:dM_diss}
\left( \frac{dM}{dt} \right) = - \frac{M(t)}{t_{\text{diss}}(t)} = - \frac{M(t)^{1 - \gamma}}{t_0},
\end{equation}
where $t_{\text{diss}}$ is the dissolution timescale that depends on the actual cluster mass and its environment, and it is assumed as a power-law function of mass, as
\begin{equation}
\label{eq:t_diss}
t_{\text{diss}} = t_0 \left( \frac{M}{M_\odot} \right)^\gamma,
\end{equation}
with $t_0$ being the dissolution parameter.

Generalising this result \citep{2021MNRAS.503.3000M}, it is common to model the mass loss simply by an expression of the form
\begin{equation}
\label{eq:dM_dt}
\dot{M} \equiv \left( \frac{dM}{dt} \right) = - \sum_i \frac{M}{\tau_i (M/M_0)^{\gamma_i}}
\end{equation}
where each term in the sum equals the current mass divided by a characteristic timescale, that in turns is proportional to the mass to some power.

When two or more terms are presented in the sum in Equation~\ref{eq:dM_dt}, there is no analytic solution for every $\gamma_i$ and $\tau_i$, instead a numerical integration must be used to find a solution. However, there is an analytical solution when we are dealing only with one term of the sum. For $\gamma \neq 0$,
\begin{equation}
\label{eq:One_term}
M(t) = M_0 \left( 1 - \gamma \frac{t}{\tau} \right)^{1/\gamma}.
\end{equation}

Our approach follows that of \citet{2021MNRAS.503.3000M} such that we first attempt a fitting to our simulations with just a single term; if this does not provide a good fit, a new fitting is attempted with two terms. 

In order to fit our mass loss curves we created a code using \texttt{emcee}\footnote{\url{https://emcee.readthedocs.io/en/stable/}} \citep{ 2013PASP..125..306F}, which is an implementation in \texttt{Python}\footnote{Python Language Reference, version 3.6. Available at http://www.python.org} of one of the Markov Chain Monte Carlo (MCMC) sampling methods, called affine-invariant ensemble samplers \citep{2010CAMCS...5...65G}.

\begin{figure}
	\includegraphics[width=\columnwidth, trim=20 10 10 0]{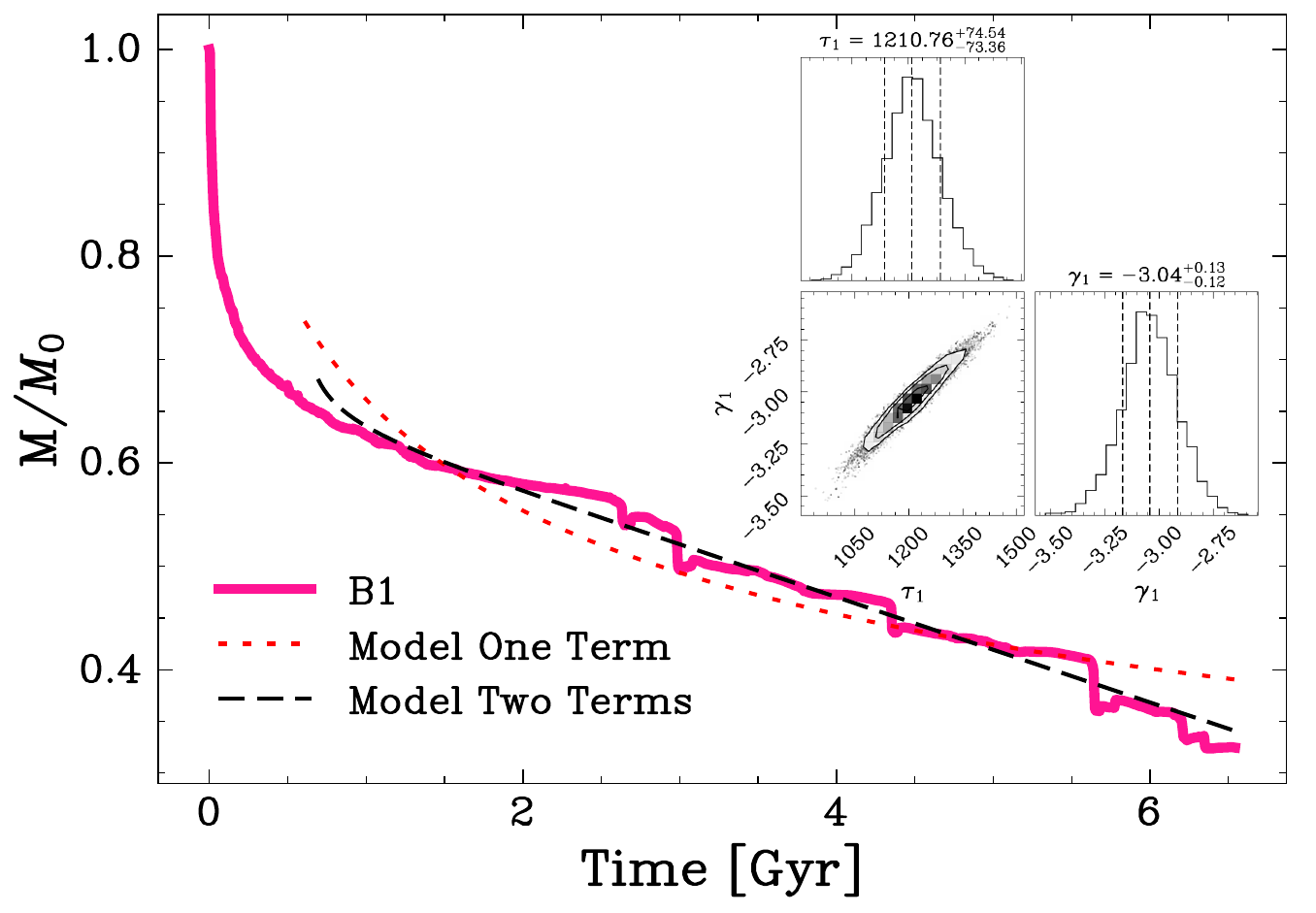}
	\centering
    \caption{
    Evolution models for GC B1: one with a single term as in eq.~\ref{eq:One_term} (dashed red line), and another with two terms in eq.~\ref{eq:dM_dt} (dotted black line). A notable variation between the models is observed, particularly towards the end of the GC's evolution. In the corner plot, the distributions of the parameters $\tau_1$ and $\gamma_1$ are displayed. The vertical dashed lines represent the median values and $1\sigma$ uncertainties of the posterior distribution for the mass loss curve model with one term.}
    \label{fig:EMCEE_MODEL}
\end{figure}

As a comparison of the fitting method to the mass loss curve of one of our simulated GCs, we used one and two terms in the sum of equation~\ref{eq:dM_dt} as shown in Figure~\ref{fig:EMCEE_MODEL}. This GC is the one labelled as ``B1'' and it was simulated with $N_0 = 32,000$ particles. In the corner plot, we can also visualise the posterior distributions of the two free parameters $\tau_1$ and $\gamma_1$ of the one-term model. From the distributions, we can see that there is a linear correlation between $\tau_1$ and $\gamma_1$. We also found that the value of $\tau_1$ is related to some kind of offset from the mass loss curve that we are modelling, and $\gamma_1$ is related to the slope of the fit. It means that if we increase the $\tau_1$ value, we will see an offset on our fit that can be compensated by changing the $\gamma_1$ value to keep the fit close to the mass loss curve.

The complete mass loss models are shown in Figures~\ref{fig:modelos_32},~\ref{fig:modelos_64} and ~\ref{fig:modelos_96} in Appendix~\ref{appendix:FittingModels}, as well as in Table~\ref{tab:params_results}, that contains the parameters found for the mass loss curve fits. 
Note how the models with two terms in the sum improve over those with a single term, which was expected due to the additional free parameters. However, in order to compare the two models and determine the most suitable for our data, we apply two commonly used information criteria. These criteria, namely the Akaike Information Criterion (AIC) and the Bayesian Information Criterion (BIC) help us strike a balance between the goodness of fit and the complexity of the model, where the preferred model will display lower AIC or BIC values. In Table \ref{tab:model_comparison} we show, for each simulation, the corresponding AIC and BIC values for both models, and in general we favour the model with lower AIC and BIC values, which in most cases corresponds to the model with two terms. Note, however, that in some cases these criteria favour the one-term model. In such cases, we opted to keep the two-terms models, because we are interested in the predicted cluster dissolution time, which depends on how good is the fit towards the end of the simulation data, which is always better for the two term model. On the other hand, the one-term model fit lies always above the data at the end of the evolution, and becomes shallower at larger times, overestimating the dissolution times.

This mass loss models are necessary in order to obtain the dissolution times of GCs that have not yet been disrupted in our simulations, and for those GCs that lived less than 12 Gyr, it is immediate to obtain their dissolution time from the simulations, which would be the time at which most of the stars in the cluster become gravitationally unbound.

\citet{2003MNRAS.340..227B} and \citet{2021MNRAS.503.3000M} define the dissolution time $t_{\text{diss}}$ as the time it takes for the remaining mass to be equal to 7\% of the initial total mass. We adopted this definition and extrapolated the fitting curves until the mass of the GC reaches 7\% of its initial value, and the obtained $t_{\text{diss}}$ values for each GC are shown in Table~\ref{tab:tdiss_results}. Note the wide range of projected dissolution times, which are much larger for clusters in galaxies A and B than for the rest of galaxies.

\begin{table}
\begin{center}
\begin{tabular}{ |l|c|c|c|c|c|c|c|c|c|c|c|c|} 
 \hline
Model & $t_{\text{diss}}$ [Gyr] & & $t_{\text{diss}}$ [Gyr] & &  $t_{\text{diss}}$ [Gyr] \\ 
\hline
\hline
         & \multicolumn{1}{c}{ Set 1 } & & \multicolumn{1}{c}{ Set 2 } & & \multicolumn{1}{c}{ Set 3 } \\
\cline{2-2} \cline{4-4} \cline{6-6}
 A1  & 31.539 & & 121.782 & & 498.170  \\ 
 B1  & 11.898 & & 25.613  & & 93.207   \\ 
 B2  & 49.424 & & 306.722 & & 1058.568 \\ 
 C1  & 7.403  & & 7.643   & & 8.944    \\ 
 D1  & 6.698  & & 8.308   & & 12.103   \\ 
 D2  & 3.926  & & 6.805   & & 8.169    \\ 
 E1  & 4.232  & & 5.518   & & 4.934    \\ 
 E2  & 3.865  & & 7.560   & & 14.620   \\ 
 \hline
\end{tabular}
\caption{\label{tab:tdiss_results} Dissolution time for our simulated GCs in the three sets of simulations.}
\end{center}
\end{table}

To quantify the mass lost by individual GCs, we proceed as follows: For GCs whose $t_{\text{diss}}$ is less than 12 Gyr, their average mass loss percentages per Gyr are computed as,
\begin{equation}
    \overline{\dot{M}}(\%) = \frac{|\Delta M|}{\Delta t}(\%) = \frac{100}{t_{\text{diss}}} \frac{|M_{t_{\text{diss}}} - M_0|}{M_0}
\end{equation}
and since we are assuming that the cluster is dissolved when it only has 7\% of its initial mass left, $M_{t_{\text{diss}}} = 7\% M_0$.

Meanwhile, for GCs whose $t_{\text{diss}}$ is longer than 12 Gyr, we decided to make a cut at this time so we are actually computing their mass loss percentages per Gyr at $t = 12$ Gyr, a value that we take as an average age for present day GCs. Hence, for such cases
\begin{equation}
    \overline{\dot{M}}(\%) = \frac{|\Delta M|}{12 \hspace{0.05cm} \text{Gyr}}(\%) = \frac{100}{12 \hspace{0.05cm} \text{Gyr}} \frac{|M_{12 \hspace{0.05cm} \text{Gyr}} - M_0|}{M_0}.
\end{equation}

\begin{figure*}
    \includegraphics[width=\textwidth]{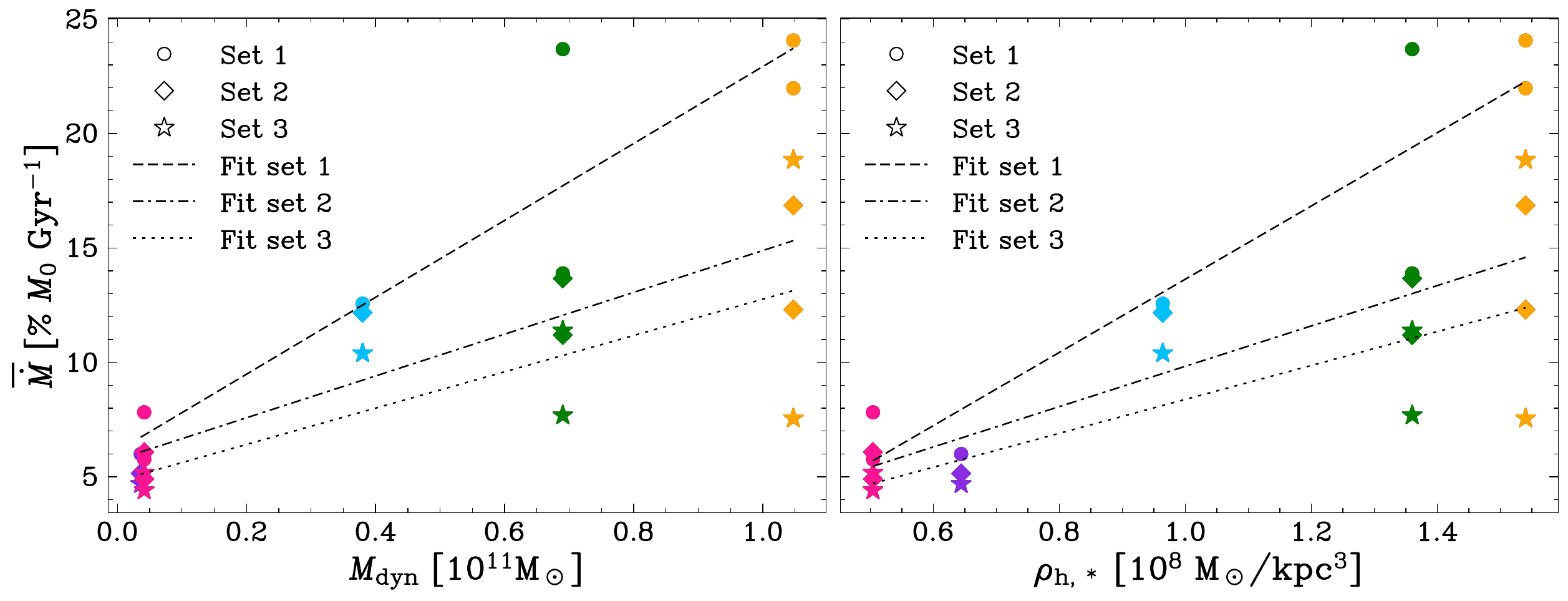}
	\centering
    \caption{Comparison of mass loss rates of GCs with respect to the dynamical mass, $M_{\rm dyn}$, of their host galaxies (left) and average total matter density within half stellar mass radius, $\rho_{\rm{h},*}$ (right). Each set of simulations is represented by distinct marker shapes (circles, diamonds, and stars) with different colours. The black curves represent the fitted models for each set. The figure shows the relationship between the percentage of mass loss rates exhibited by GCs and the corresponding host galaxy mass and density, providing valuable information about how effective is the disruption of GCs in different environments.}
    \label{fig:MassLossPercentage}
\end{figure*}

Figure~\ref{fig:MassLossPercentage} shows the computed mass-loss rates (in percentage of the initial cluster mass) for our three sets of $N$-body models, as a function of the host galaxy dynamical mass, $M_{\rm dyn}$, and mass density, $\rho_{\rm{h},*}$. We also applied a linear regression to each simulation set (black lines), which is able to capture the overall behaviour of cluster mass loss rates as a function of $M_{\rm dyn}$ and $\rho_{\rm{h},*}$.

Our findings provide compelling evidence that the dissolution of GCs is strongly influenced by the mass and density of their host galaxy. Specifically, in our sample, we observe that GCs in more massive galaxies experience accelerated dissolution, while GCs in the least massive galaxies, namely galaxies A and B, exhibit a less pronounced dissolution process. This observation highlights the crucial role played by the mass of the galactic system in determining the evolution of GCs.

\subsection{Evolving CG Systems}
\label{sec:GC_systems}

Our research focuses on understanding how GC Systems (GCSs) evolve within their host galaxies and how much mass is lost from GCs, eventually becoming part of the host galaxy's stellar population. This analysis can provide context for the specific frequency problem mentioned in the introduction. To accomplish this, here we create a synthetic system of GCs for each galaxy, and then evolve each cluster using the mass loss rates obtained from our GC simulations. Then, we measure the overall mass loss of the GCS. One important thing to consider is that we are taking the tidal tensors we currently have for each galaxy as a general behaviour of the galaxy's tidal field.

We generate a system of GCs using a power-law distribution as the initial GC mass function (IGCMF), that follows the form $dN/dM \propto M^{-2}$, where $dN/dM$ represents the number of GCs per unit mass range and $M$ is the GC mass. This functional form is proposed based on empirical derivations and observations of young star clusters \citep{1999ApJ...527L..81Z, 2003MNRAS.343.1285D, 2003A&A...397..473B, 2007ApJ...663..844M, 2008AJ....135..823D}. The power-law distribution describes the distribution of GC masses in the population, where the number of GCs decreases as the mass increases. This power-law function allows us to simulate the statistical distribution of GC masses within the synthetic GC system.

\begin{figure}
    \includegraphics[width=\columnwidth, trim=20 10 10 0]{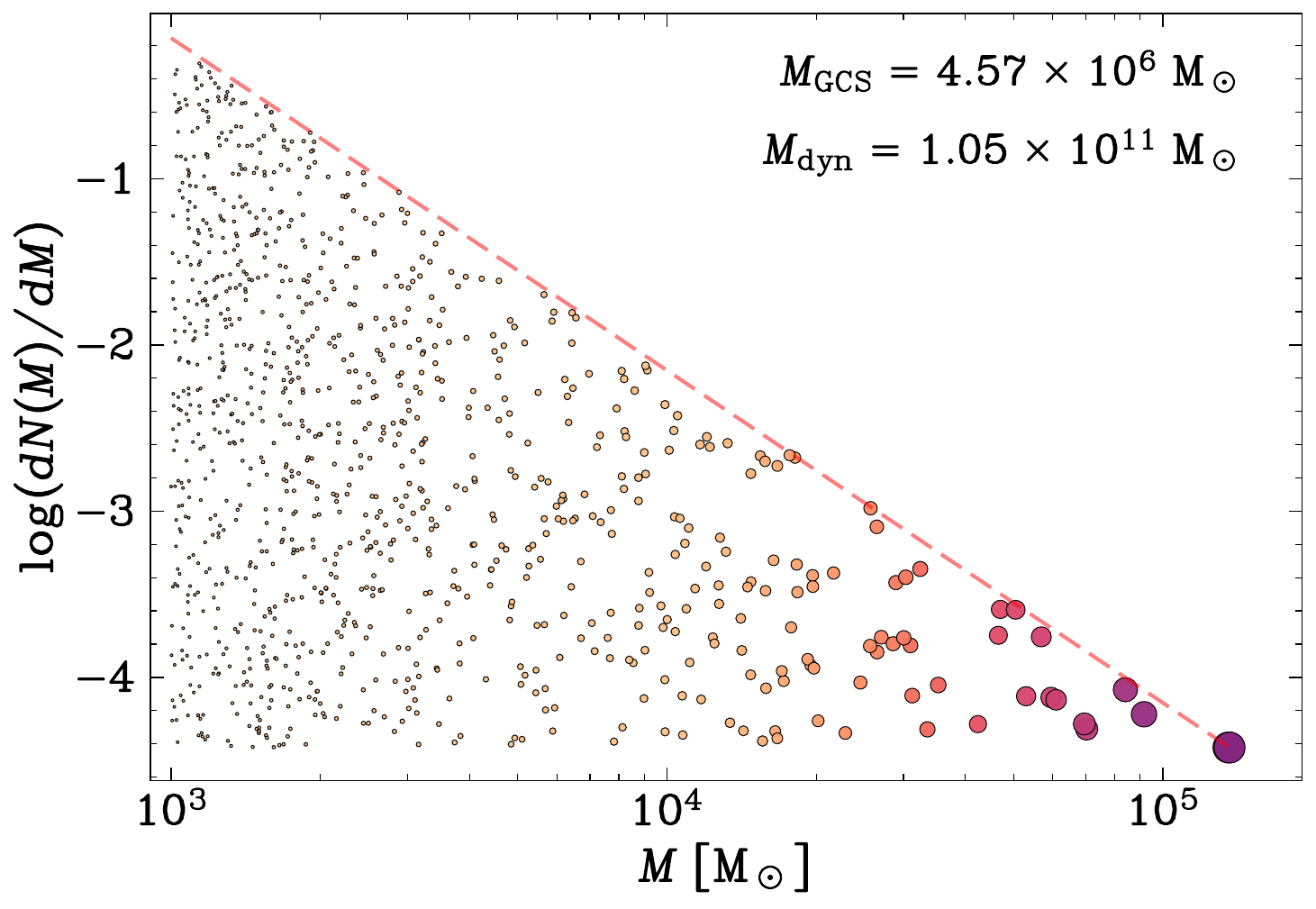}
	\centering
    \caption{The GC systems are generated using a power-law distribution as the IGCMF, following the form $dN/dM \propto M^{-2}$. The figure demonstrates the power-law distribution of GC masses within the system. Each dot represents an individual GC, and the dot's size corresponds to the mass of that cluster. Thus, the bigger the dot, the more massive that cluster is in the GC system, highlighting the abundance of lower-mass GCs and the decreasing frequency of higher-mass ones.}
    \label{fig:IGCMF_orange}
\end{figure}

To generate the GCS, we use a stochastic random sampling method. It involves generating a set of random numbers between 0 and 1 for each GC in the system. These random numbers are then used to assign masses to the GCs based on the probability distribution defined by the power-law function. To create the GC population in a given galaxy, several steps were taken within our modelling framework. First, the minimum mass value was fixed at $M_\text{min}= 1\times 10^3$ M$_{\odot}$, serving as the lower limit for the masses of the sampled clusters. Next, we compute the initial mass of the GCS, $M_{\text{GCS},0}$, using the relation proposed by \cite{2023MNRAS.518.2453D} that, after the time evolution of their GCSs, reproduces the observed relationship between $M_{\text{GCS}}$ and the mass of the galaxy \citep{2015ApJ...806...36H}.
Doppel's relation returns the mass of the GCS at the time of infall, providing a more accurate approximation of the system's initial mass. This relation is given by:
\begin{equation}
    M_{\text{GCS,inf}} = a_{\text{inf}} M_{\text{halo}, \text{inf}}^{b_{\text{inf}}}.
\end{equation}
Furthermore, our study focuses on the metal-poor population of GCs, also called blue component, in dwarf galaxies. Therefore, we use the $a_{\text{inf}}$ and $b_{\text{inf}}$ values reported in \cite{2023MNRAS.518.2453D} for metal-poor GCs, i.e., $a_{\text{inf}} = 7.3 \times 10^{-5}$ and $b_{\text{inf}} = 0.98$. 

Finally, to determine the maximum mass for our sampling of the IGCMF, we use equation 12 from \cite{2010ApJ...718.1266M},
\begin{equation}
\label{eq:Mmax}
    M_{{\rm GCS},0} = M_{\text{max}} \ln{\frac{ M_{\text{max}} }{ M_{\text{min}} }}
\end{equation}
where $M_{\text{max}}$ represents the maximum mass that can be assigned to an individual cluster.

By applying this procedure, we obtained the values for $M_{\text{max}}$ and $M_{\text{GCS},0}$ that correspond to the mass $M_{\text{dyn}}$ for each of our five galaxies, as shown in Table~\ref{tab:Mmax_MGCS}. With this information, we were able to sample the IGCMF and generate a GCS for each galaxy. As an example, Figure~\ref{fig:IGCMF_orange} illustrates the generated GCS for galaxy E, with the $x$-axis representing the cluster mass and the $y$-axis indicating the number of clusters per unit mass range.

\begin{figure*}
    \includegraphics[width=\textwidth]{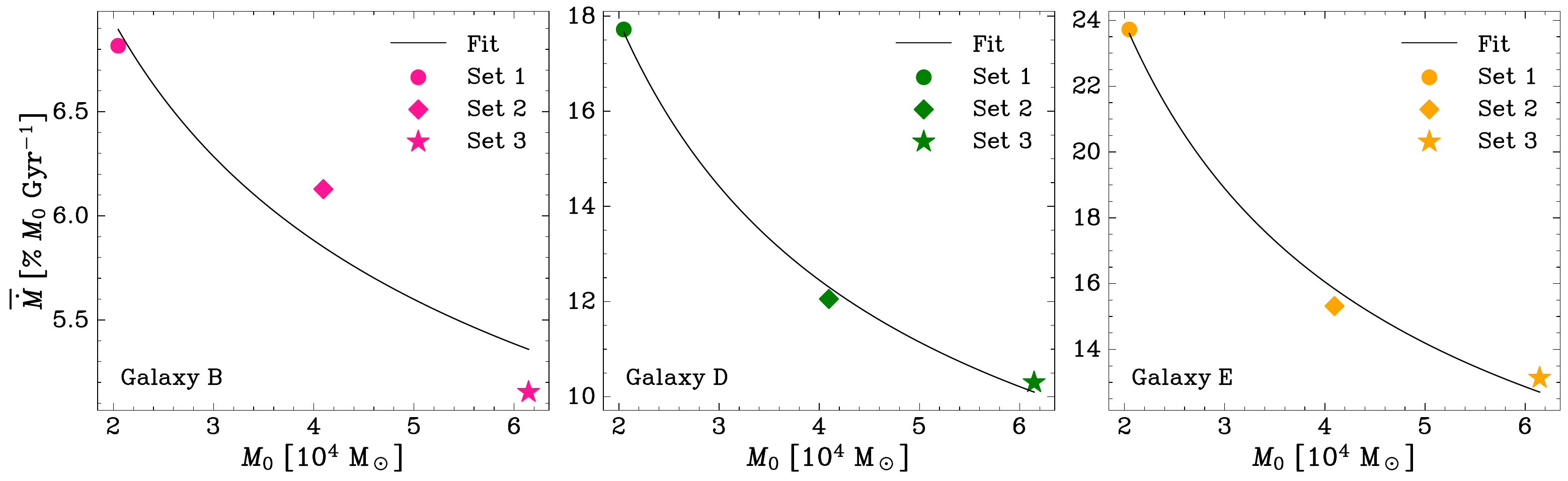}
	\centering
    \caption{Models of the fitted mass loss results for the GCs simulations as a function of the initial mass, $M_0$. Each figure represents a different galaxy and displays the fitted power-law curve (solid line) along with the mass loss rates obtained from the linear regression for each set of simulations (represented by circles, diamonds and stars). These models allow us to extrapolate the mass loss percentages that more massive GCs would exhibit in different galaxies.}
    \label{fig:Percentages_extrapolation}
\end{figure*}

Next, we proceed to evolve each system, for which we know that the evolution of an individual cluster depends on its initial mass $M_0$. To assign a mass loss rate to each cluster we use the results from our $N$-body models as summarised in Figure \ref{fig:MassLossPercentage}, where the three linear fits (black lines) provide mass loss rates for clusters with the same initial density but different initial mass. Hence, for each galaxy we take this mass loss rate as a function of $M_0$ and fit the data using a power law in $M_0$. Figure~\ref{fig:Percentages_extrapolation} shows examples of this procedure for three of the galaxies, where the fitting curve is used to draw mass loss rates for individual clusters according to their initial mass $M_0$, therefore, the entire system can be evolved in time. In this sense, the mass evolution of our GC systems was calibrated using the results from our $N$-body models. 

\begin{figure*}
    \includegraphics[width=\textwidth, trim=20 10 10 0]{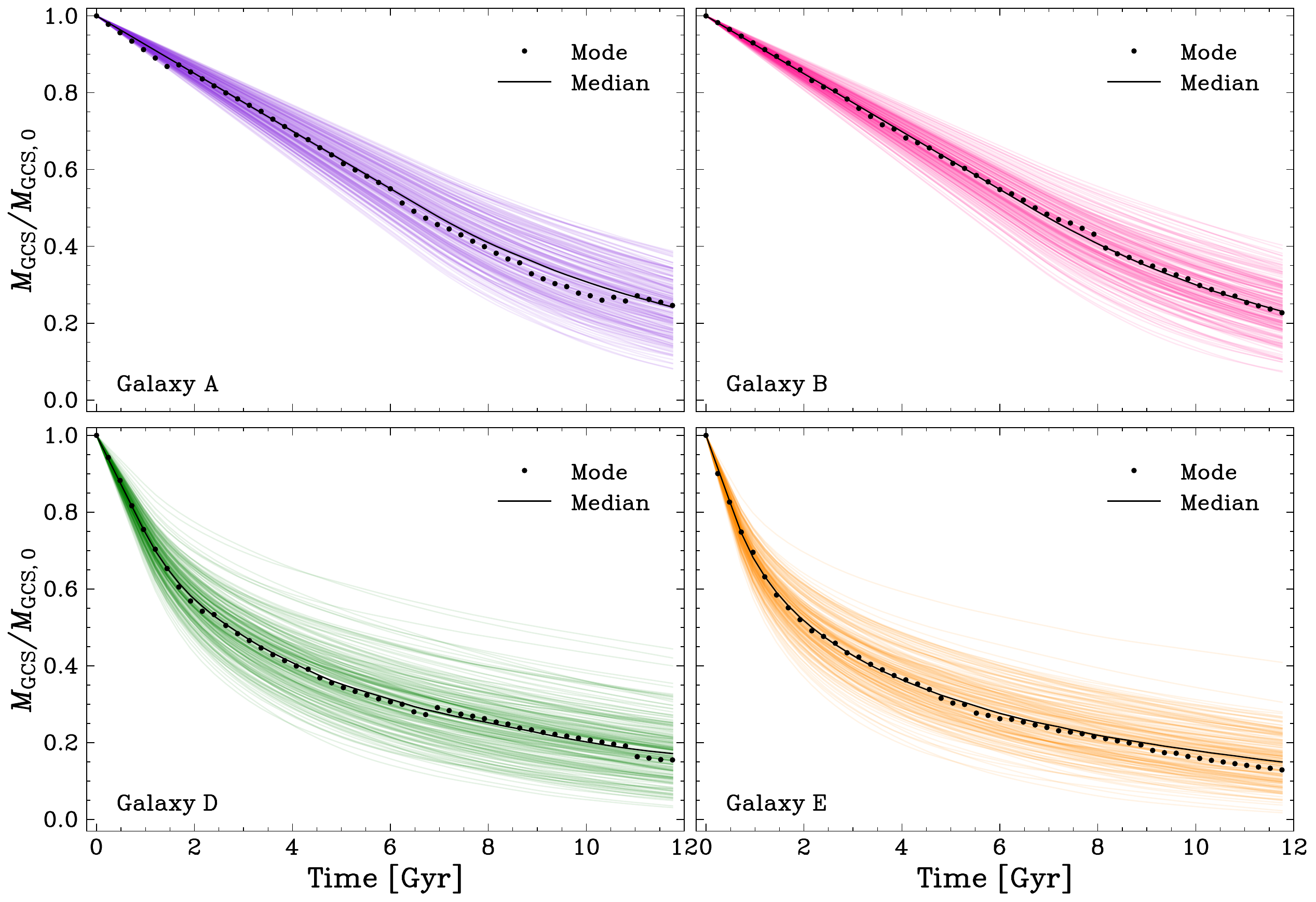}
	\centering
    \caption{Evolution of total mass in GC systems. Each subplot represents the distribution of mass for a set of GC systems within a galaxy, and the curves within each subplot depict the mass evolution for each individual GC system. The colours correspond to different galaxies, here we are showing galaxies A, B, D, and E. The black lines and markers represent statistical measures: mode (black dots) and median (solid black lines).}
    \label{fig:GCS_Evolution_Distribution_Statistics}
\end{figure*}

In Figure~\ref{fig:GCS_Evolution_Distribution_Statistics} we present the temporal evolution of multiple GCSs, where we evolve a sufficiently large number of systems for each galaxy to account for the inherent stochasticity in the sampling of the IGCMF. Each system and its evolution are unique due to this stochastic process. Additionally, it is worth noting that the curves located more spread up in the distributions represent less probable cases, while the regions with higher density of lines indicate the most probable evolution of our systems. As time progresses, the ensemble of curves tends to disperse, reflecting the increasing uncertainty in the evolution of the GCS.

\begin{figure}
	\includegraphics[width=\columnwidth, trim=20 10 10 0]{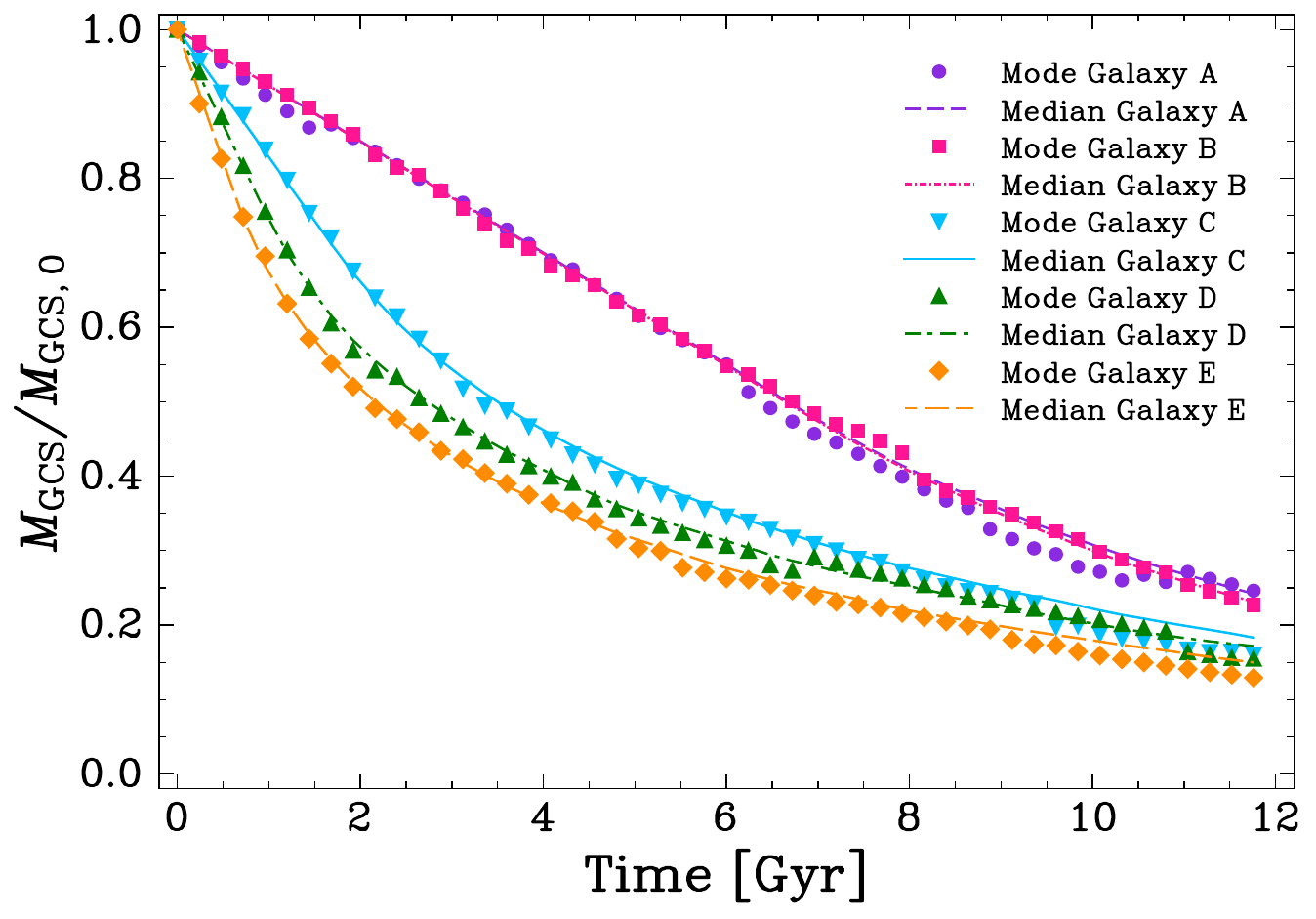}
	\centering
    \caption{Comparative analysis of mass loss statistics in different GC systems. The plot shows the mode (dots with different symbols) and median (dashed lines with different styles) curves representing the evolution of mass over time for various GC systems in the five different host galaxies. The galaxies, denoted as 'Galaxy A' to 'Galaxy E,' are indicated by different coloured curves.}
    \label{fig:GCS_Evolution_OnlyStatistics}
\end{figure}

To obtain information on the overall behaviour of these ensembles we applied two statistical measures: mode and median; these measures allow us to capture different aspects of the GCS evolution within our five galaxies. In Figure~\ref{fig:GCS_Evolution_OnlyStatistics} we present the comparison of these statistical results. The mode (dotted lines) represents the most frequently observed value in the population of GCSs at each time step. This measure helps to identify the regions of higher probability in the evolution of the GCS. The solid lines correspond to the median, which represents the central tendency of the data distribution. It provides insight into the typical behaviour of the GCS populations. It is worth noting that the median differs from the mode, indicating a slightly skewed distribution of curves and better reflecting the evolution of each GCS. Hence, the median and mode provide a comprehensive understanding of the dynamics of GCSs and the efficiency of disruption mechanisms operating in the host galaxies.

\begin{table}
\begin{center}
\begin{tabular}{ |l|c|c|c|c|} 
 \hline
Galaxy & $M_{\text{dyn}}$ [$M_\odot$] & $M_{\text{GCS},0}$ [$M_\odot$] & $M_{\text{max}}$ [$M_\odot$] & $\rho_{\rm{h},*}$ [$M_\odot/\rm{kpc}^3$] \\ 
\hline
\hline
A    & 3.606e+09  &  1.696e+05   & 4.463e+04    & 6.44e+07 \\ 
B    & 4.147e+09  &  1.944e+05   & 4.975e+04    & 5.04e+07 \\ 
C    & 3.799e+10  &  1.704e+06   & 2.989e+05    & 9.64e+07 \\ 
D    & 6.901e+10  &  3.058e+06   & 4.932e+05    & 1.36e+08 \\ 
E    & 1.047e+11  &  4.603e+06   & 7.023e+05    & 1.54e+08 \\ 
 \hline
\end{tabular}
\caption{\label{tab:Mmax_MGCS} Comparison of dynamical Mass ($M_{\text{dyn}}$), Globular Cluster System Mass ($M_{\text{GCS},0}$), maximum Mass ($M_{\text{max}}$) and average total matter density within half stellar mass radius ($\rho_{\rm{h},*}$) at $z\sim3$ for galaxies A, B, C, D and E.}
\end{center}
\end{table}

\section{DISCUSSION}
\label{section:Discussion}

The GC specific frequency has been extensively studied in the literature, covering a wide range of galaxy luminosities and environments \citep[e.g.][]{1981AJ.....86.1627H, Peng2008, 2010MNRAS.406.1967G, 2012A&A...537A...3M, 2017ApJ...849....6A, 2019ApJ...875..156L}. This property can also be expressed as the number of GCs ($N_{\rm GC}$) per stellar mass of the host galaxy \citep{Peng2008}, or as the stellar mass in the system of GCs ($M_{\rm GCS}$) per dynamical mass of the host \citep{Harris2013}. The values of these quantities are distributed in a similar way: they are high for dwarfs and giant ellipticals, with a minimum at intermediate values of $M_V$, $M_*$, and $M_{\rm dyn}$.

As already mentioned in Section \ref{section:Introduction}, a plausible explanation for this trend in the specific frequency relies on the efficiency of galaxies to form GCs, followed by a diminished field star formation that affected mainly dwarfs and high mass galaxies. However, other mechanisms are known to shape the properties of GC systems. It is well known that GC dissolution and erosion depend on the galactic environment \citep[e.g.][]{1997ApJ...474..223G, 2003MNRAS.340..227B, 2005A&A...429..173L, 2009ApJ...706...67R, Rieder2013}, and as such, they have the potential to be important contributors in setting the observed trend in specific frequency and specific mass.

Following this direction, \citet{2014A&A...565L...6M} explained the U-shaped $S_N$ as an effect of tidal erosion. They studied the relation between GC survival fractions ($f_s$) and the three-dimensional mass density within the half-light radius of their host galaxy. Their methodology consists of spherical isolated galactic models, where GCs are represented as single particles, for which mass loss mechanisms are taken into account using semi-analytical prescriptions. They found that there is a high erosion of GCs at higher galactic densities, which may explain the relation between present-day GC specific frequency and host galaxy luminosity. In other words, GC disruption can be important for the relation between $S_N$ and $M_V$.

Adopting a different approach to test the effects of GC mass loss mechanisms, we selected time-dependent dwarf galaxies from a cosmological simulation and combined them with GC {\it N}-body models. This has the advantage of accounting for a complete view of GC evolution, including not only the galactic influence (encoded in the galactic tidal field) but also the internal evolution, capturing details such as stellar evolution, binary stars formation, two-body relaxation, and evaporation.

Figure~\ref{fig:MassLossPercentage} summarises the results from our simulations, where we show the mean GC mass loss per Gyr, $\overline{\dot{M}}$, as a function of the dynamical mass of the host galaxy. Our data show a clear correlation between $\overline{\dot{M}}$ and $M_{\rm dyn}$, which we quantify by applying a linear regression for the three sets of simulations. The best-fitting parameters obtained for the mass loss rates of GCs with respect to the dynamical mass, $M_{\rm dyn}$ are as follows:

\begin{itemize}
\item Set 1: $m_1 = 1.68 \times 10^{-10}$, $b_1 = 6.12$
\item Set 2: $m_2 = 9.14 \times 10^{-11}$, $b_2 = 5.75$
\item Set 3: $m_3 = 7.94 \times 10^{-11}$, $b_3 = 4.82$
\end{itemize}

And the best-fitting parameters obtained for the mass loss rates of GCs with respect to the average total matter density within half stellar mass radius, $\rho_{\rm{h},*}$ are as follows:

\begin{itemize}
\item Set 1: $m_1 = 1.60 \times 10^{-7}$, $b_1 = -2.36$
\item Set 2: $m_2 = 8.82 \times 10^{-8}$, $b_2 = 1.01$
\item Set 3: $m_3 = 7.41 \times 10^{-8}$, $b_3 = 0.97$
\end{itemize}

We find that the mass loss rate of GCs is lowest in our less massive galaxies and increases for the more massive ones. This relation implies that the total mass in a GC system, $M_{\rm GCS}$, is diminished with less efficiency in less massive dwarfs, which would have an impact on relevant quantities. For example, the specific mass, defined as $S_M=100~M_{\rm GCS}/M_{\rm dyn}$, is directly affected by GC mass loss rates that depend on $M_{\rm dyn}$, i.e., the disruption mechanisms favour high $S_M$ values in galaxies at the low-mass end; on the other hand, for larger $M_{\rm dyn}$ values, $S_M$ is further reduced by a more efficient GC mass loss, and this trend is preserved across the range of galactic masses explored in this work.

Here it is worth mentioning that although we explore a single value for the initial cluster density, star clusters form within a range of densities, and this property has important implications for their evolution. In general, the two-body relaxation time scale and how the cluster responds to the external tidal field depend on the cluster’s density profile. In a recent paper, \citet{Gieles2023} use $N$-body simulations and show that higher density star clusters experience systematically lower mass loss rates, hence, displaying larger dissolution times. With this results in mind, and under the assumption that the initial density distribution for clusters in each GCS does not varies significantly among our galaxies, we would expect no significant changes in the trends we have found. Particularly, models with initial density larger than the one used here would result in lower $\overline{\dot{M}}$ values in Figure \ref{fig:MassLossPercentage}, and larger remnant mass $M_{\rm GCS}$ for the GCSs.

Regarding the second specific frequency problem, where the ratios of metal-poor GCs to metal-poor stars are even higher, \citet{Larsen2012,Larsen2014} estimated the fractions of metal-poor stars that belong to the most metal-poor GCs in the Fornax, WLM, and IKN dwarf galaxies. They were able to put constraints to the initial mass of such GCs, and on the amount of mass lost from metal-poor star clusters, which is appreciably less compared to clusters in the MW stellar halo.

In Section \ref{sec:GC_systems} we addressed this second specific frequency problem by generating and assigning a GC system to each one of our five galaxies. Then, by applying mass loss rates motivated by our $N$-body models, we evolved the mass of each individual GC in the system. With this approach, we obtained the mass evolution for every GC system, and in particular we can compare their remaining mass after 12 Gyr. Figure \ref{fig:GCS_Evolution_Distribution_Statistics} shows the mass evolution of these GC systems as a function of time, where the mode and median curves describe the overall behaviour that the GC systems have in each galaxy. Here, the early mass loss of the system is dominated by the disruption of the less massive GCs, and the disruption of clusters proceeds towards the end until the remaining system consists only of the few initially more massive GCs that managed to survive. 

Meanwhile, Figure \ref{fig:GCS_Evolution_OnlyStatistics} is a comparison of the evolution of GC systems in the five dwarf galaxies. Here, we notice a few clear differences in the behaviour of the curves: First, right after the GC systems start to evolve, the slopes of the curves are stepper for the more massive galaxies, which reflects the results from our $N$-body models (see Fig. \ref{fig:MassLossPercentage}), where the mass loss of star clusters is more efficient in our more massive galaxies. Second, as the evolution proceeds, the Galaxy C, Galaxy D, and Galaxy E curves become shallower; this is because the less massive clusters have been completely disrupted and the remaining ones experience lower mass loss rates, as shown in Figure \ref{fig:Percentages_extrapolation}. Meanwhile, since the less massive clusters in the two less massive galaxies are not disrupted as fast, the pink and purple curves do not change their slope until later times.

Finally, Figure \ref{fig:GCS_Evolution_OnlyStatistics} shows that the curves get closer towards the end of the 12$~$Gyr evolution, where the systems contain only the initially most massive clusters, which for each galaxy are the ones with the lowest mass loss rates since their dissolution times scale with their initial mass. Moreover, this is combined with the fact that more massive galaxies have a larger mass budget to form star clusters, i.e., a larger number of massive clusters, and larger maximum mass for clusters in the system (according to equation \ref{eq:Mmax}); this allows Galaxy C, Galaxy D, and Galaxy E curves in Figure \ref{fig:GCS_Evolution_OnlyStatistics} to change their slopes significantly. Nonetheless, at the end of the evolution, we find a clear correlation between the remnant mass fraction of the GC system and the galaxy mass, being larger for the less massive galaxies, i.e., the $M_{\rm GCS}/M_{\rm GCS,0}$ values at 12~Gyr range from $\sim$12\% for the most massive galaxy up to $\sim$25\% for the least massive galaxy. 

Moreover, we found that the remaining $M_{\rm GCS}$ fraction in any of our five dwarfs is appreciably larger compared to the MW, where the GC system has a mass of $\sim$2\% the mass of the stellar halo \citep{Kruijssen_Portegies_2009,Larsen2012}. This result is in good agreement with observations by \citet{Larsen2012,Larsen2014} and \citet{deBoer2016}, where large fractions of metal-poor stars in some dwarfs belong to GCs, which in our case is due to the inability of the galactic tidal field to disrupt GCs as fast as in more massive galaxies, like the MW.

\section{Conclusions}
\label{section:Conclusions}

We have studied the disruption of GCs in the context of the two specific frequency problems, for which we made use of $N$-body models and time-dependent dwarf galaxies from a high-resolution cosmological simulation.

From the large body of literature devoted to quantify and understand the specific frequency of GCs, we have learned that this property can be modified and re-shaped by different processes: through quenching mechanisms that diminished field SF after early starbursts of GC formation \citep[e.g.][]{Peng2008,Harris2013}; through environmentally driven SF peaks that enhanced the number of formed GCs in dwarfs located in galaxy clusters \citep[e.g.][]{Mistani2016}; or the depletion of GCs by disruption mechanisms, whose efficiency is sensitive to the mass and density of the host galaxy \citep[e.g.][]{Murali1997,2003MNRAS.340..227B,2005A&A...429..173L,2014A&A...565L...6M}. It is likely that the observed GC specific frequency is the result of a combination of these processes, each one with an importance that depends on the range of galactic masses and environments.

Our goal was to explore the anomalously large GC specific frequencies in dwarf galaxies as a possible consequence of cluster evolution; for this purpose, we quantified the mass lost by individual GC and by GC systems. We found a clear correlation of GC mass loss rates with the host's dynamical mass (see Figure~\ref{fig:MassLossPercentage}) implying that the more massive the host galaxy is, the more effective its globular clusters disrupt. Such a trend in disruption efficiency is consistent with a scenario where cluster mass loss mechanisms play an important role in shaping the specific frequency, along with GC formation and galaxy evolution. Moreover, when we compute the mass loss rates from our $N$-body clusters as a function of the galaxy density we find a positive correlation, in agreement with results by \citet{Murali1997} and \citet{2014A&A...565L...6M}.

We also addressed the second specific frequency problem from a cluster disruption point of view. For each of our selected galaxies we constructed GC systems, whose properties ($M_{\rm GCS,0}$ and $M_{\rm max}$) depend on the host's dynamical mass; then, we extrapolated the mass loss rates from our $N$-body models to a wider range of initial cluster masses, which allowed us to evolve each GC system. In Figure \ref{fig:GCS_Evolution_OnlyStatistics} we compared the fractions of mass that after 12 Gyr still belong to the systems in each dwarf, resulting in a fraction that is larger for the less massive galaxies, and decreases for increasing galactic masses. Meanwhile, when we compare these results with the case of the MW we found that the mass left in GCs is significantly larger for any of our dwarfs. This result is consistent with the interpretation of the second specific frequency problem in dwarf galaxies as a consequence of GCs that have not lost enough mass, compared with clusters in the MW's halo.
 
In general, the high specific frequency of GCs in dwarf galaxies cannot be fully explained from the perspective of how efficient GC and field star formation was, since it is also determined by how efficient cluster disruption is in these types of environments according to current models and constraints on GC evolution. The advantage of this work is that we are modelling GCs as true $N$-body systems, embedded in time-dependent cosmologically-motivated galaxies. However, we believe that it can be improved by using simulations with higher initial numbers of particles, which would allow to infer more accurate mass loss rates for GCs at the high-mass end. Moreover, in an upcoming paper we seek to extend our study to a broader range of galactic masses, where GC disruption efficiencies might follow different trends and have variable influences on the observed U-shaped specific frequency (Moreno-Hilario et al., in preparation).

\section*{Acknowledgements}

We thank the anonymous referee for a careful review and insightful suggestions that contributed to improve this work. EMH and LAMM acknowledge support from DGAPA-PAPIIT IA101520 and IA104022 grants. We acknowledge DGTIC-UNAM for providing HPC resources through the project LANCAD-UNAM-DGTIC-369. EMH is supported by a CONACYT scholarship. SOS acknowledges the FAPESP PhD fellowship number 2018/22044-3. APV, EMH, and SOS acknowledge the DGAPA–PAPIIT grant IA103122.

{\it{Facilities}}: Supercomputers Miztli and ATOCATL.

{\it{Software}}: 
\texttt{numpy} \citep{van-der-Walt:2011aa},
\texttt{matplotlib} \citep{Hunter:2007aa},
\texttt{pandas} \citep{mckinney-proc-scipy-2010, reback2020pandas},
\texttt{SciPy} \citep{Virtanen:2020aa},
\texttt{corner} \citep{Foreman-Mackey:2016aa},
\texttt{emcee} \citep{2013PASP..125..306F}.

\section*{Data Availability}

The data that support the findings of this study are available from the corresponding author, upon reasonable request.




\bibliographystyle{mnras}
\bibliography{example} 




\appendix

\section{Modelling the Evolution of Globular Cluster Masses}
\label{appendix:FittingModels}

In Section~\ref{subsection:MassLossModels}, we discussed the modelling of GC mass loss and presented an illustrative case. Here, we provide further details regarding the models employed and the fitted parameters.

For our analysis, we fitted mass loss models using only one term in eq.~\ref{eq:dM_dt}, and after careful examination, we found that models with two terms provided better fits to the GCs simulations. These models captured more accurately the GC mass loss compared to models with a single term, as obtained from our analysis, applying the AIC and BIC information criteria. It is worth noting that the fitting procedure excluded the initial 300 - 400 Myr of GC evolution. This decision was made because the smooth nature of the fitting curves in this early stage of evolution made it challenging to capture that specific part of the simulations accurately. Our primary focus was to ensure a robust fit to the later stages of GC evolution, as this is crucial for accurately estimating the dissolution times. By employing models with two terms and excluding the initial phase, we obtained fitted parameters that best represented the observed GC mass loss. These parameters allowed us to estimate the dissolution times and study the long-term evolution of GCs in our analysis.

\begin{figure*}
    \includegraphics[width=\textwidth, trim=20 10 10 0]{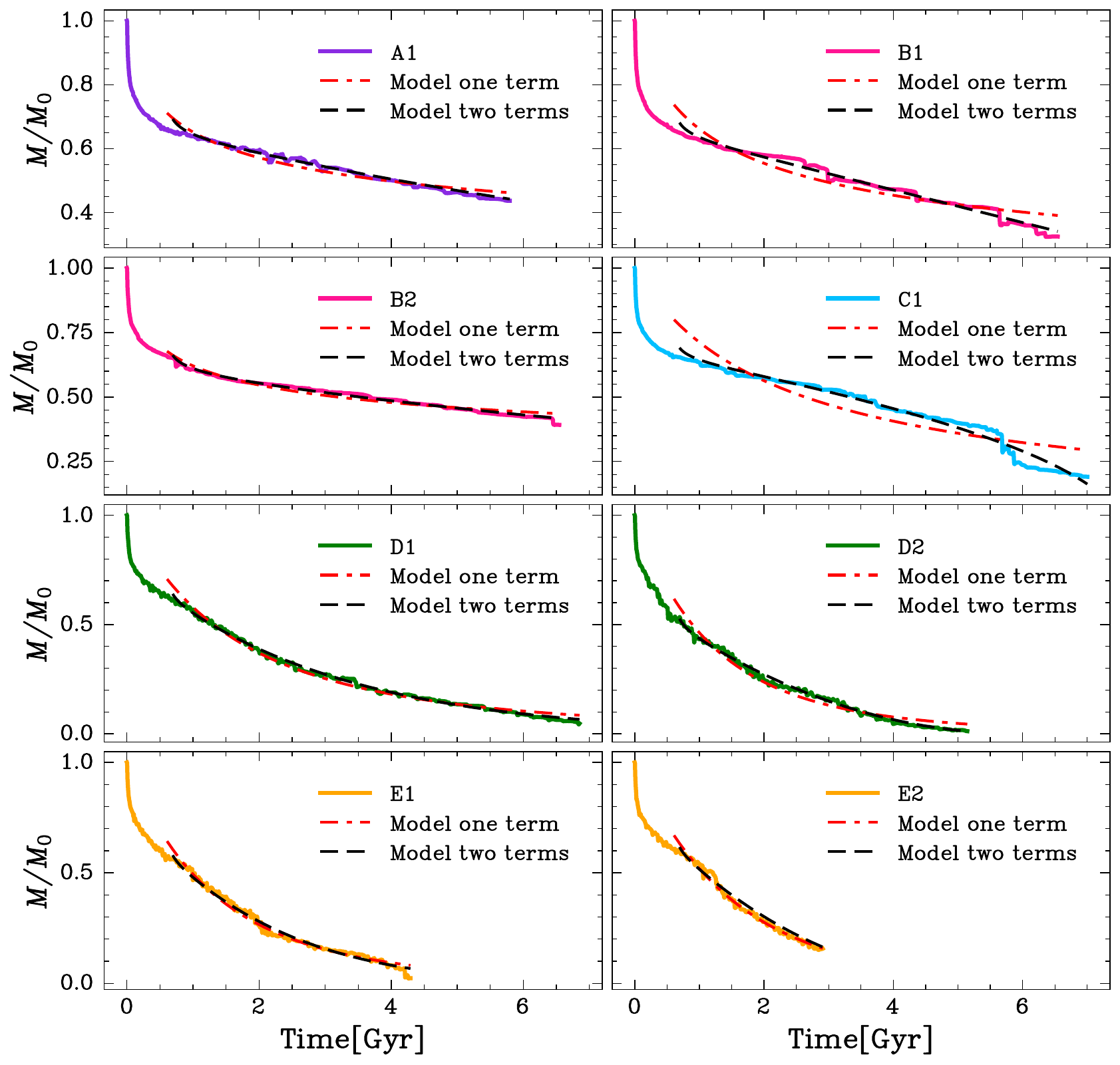}
	\centering
    \caption{Models with one term (dashed-dot lines in red) and two terms (dashed lines in black) for the evolution of mass loss in the simulations of Set 1. Each subplot represents a different GC, labeled accordingly.}
    \label{fig:modelos_32}
\end{figure*}

\begin{figure*}
	\includegraphics[width=\textwidth, trim=20 10 10 0]{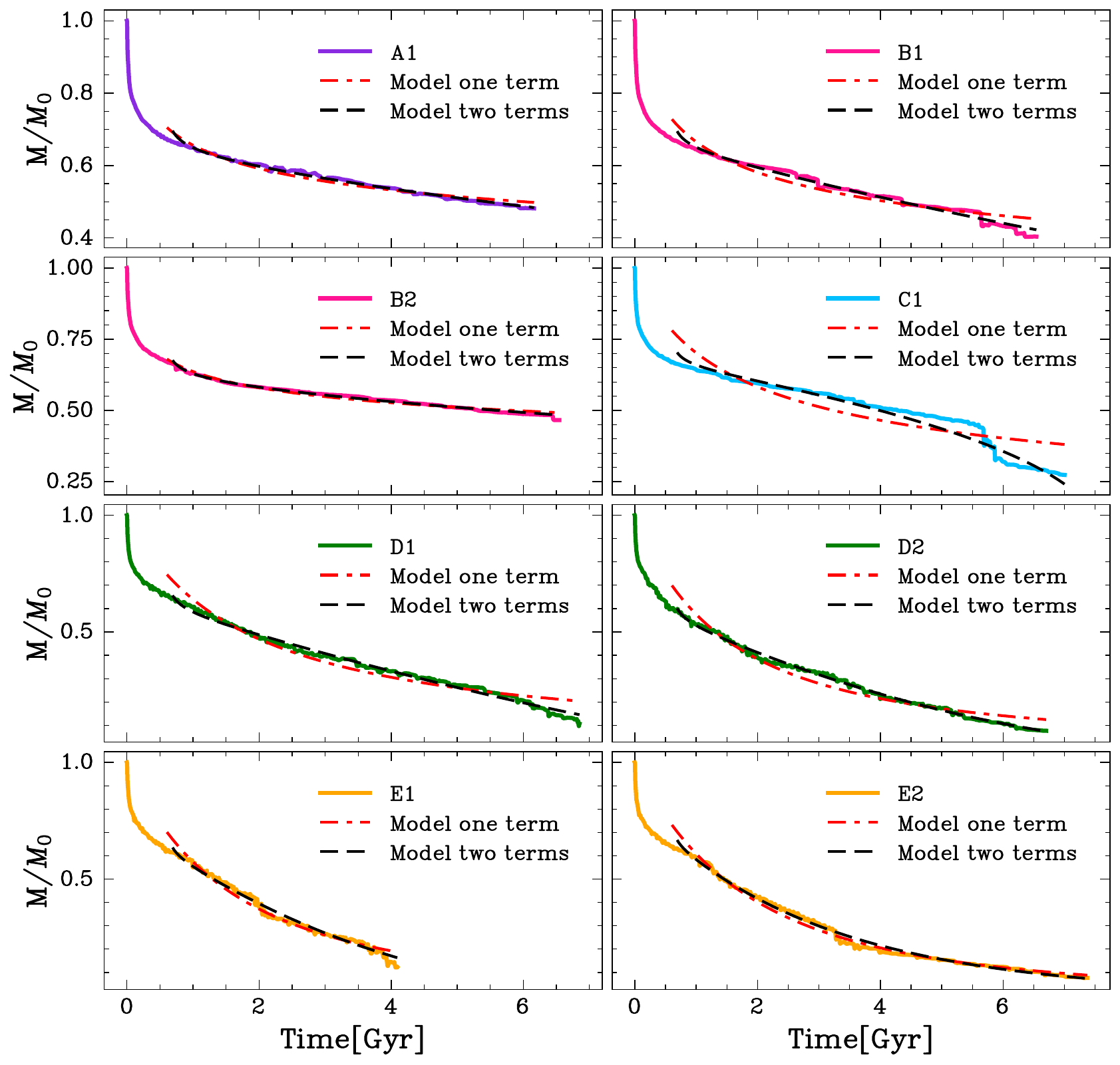}
	\centering
    \caption{Models with one term (dashed-dot lines in red) and two terms (dashed lines in black) for the evolution of mass loss in the simulations of Set 2. Each subplot represents a different GC, labeled accordingly.}
    \label{fig:modelos_64}
\end{figure*}

\begin{figure*}
	\includegraphics[width=\textwidth, trim=20 10 10 0]{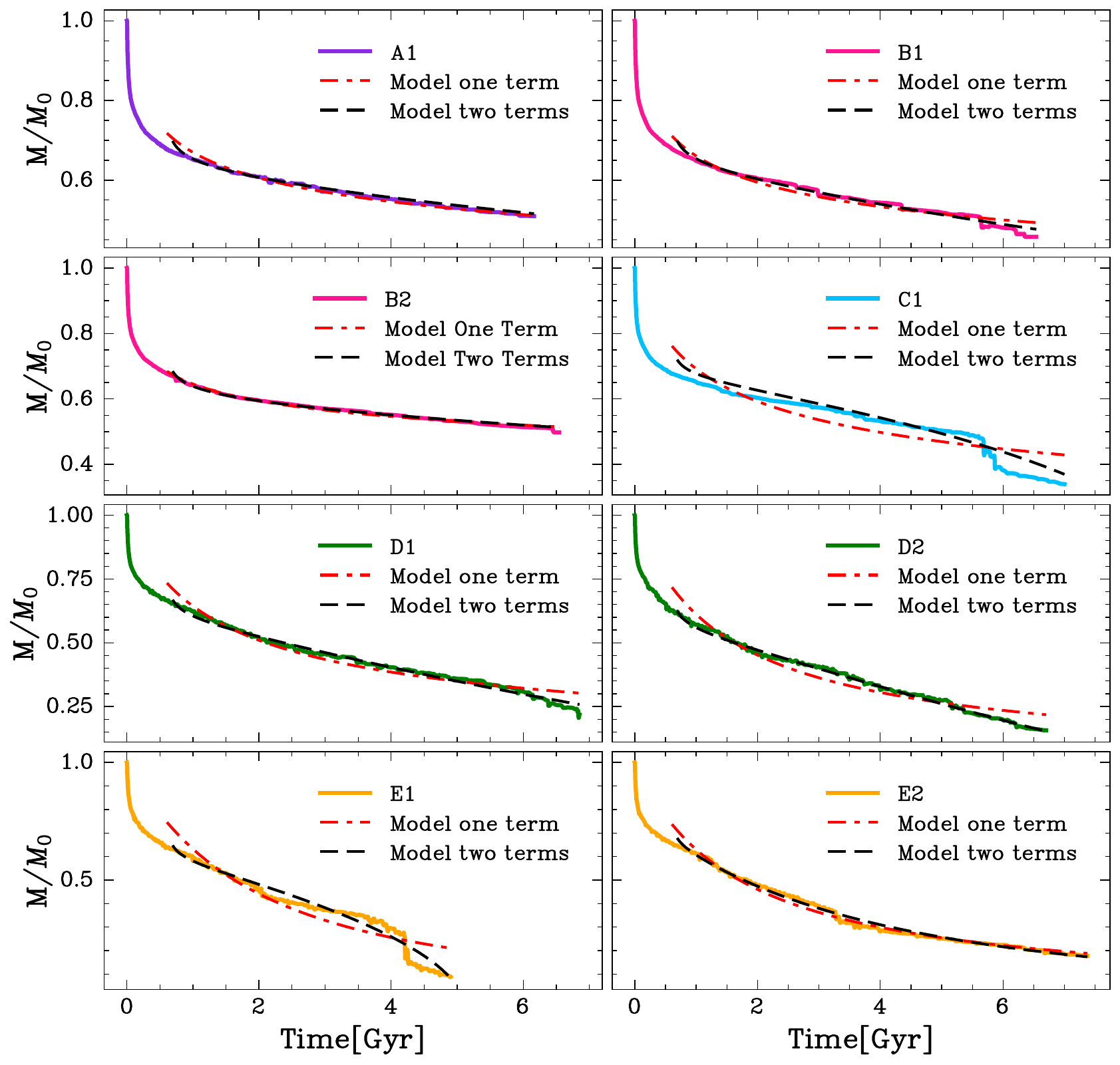}
	\centering
    \caption{Models with one term (dashed-dot lines in red) and two terms (dashed lines in black) for the evolution of mass loss in the simulations of Set 3. Each subplot represents a different GC, labeled accordingly.}
    \label{fig:modelos_96}
\end{figure*}

\begin{table*}
\begin{center}
\begin{tabular}{ |l|c|c|c|c|c|c|c|c|c|c|c|c|c|c|c|c|c|c|c|c|c|} 
 \hline
   & &  \multicolumn{2}{c}{ One-Term Model }  & & \multicolumn{4}{c}{ Two-Terms Model } \\
GC & & $\tau_1$ & $\gamma_1$ & & $\tau_1$ & $\gamma_1$ & $\tau_2$ & $\gamma_2$ \\ 
\hline
\hline
& &  \multicolumn{7}{c}{ Set 1 } \\
\cline{3-9} \\
 A1 & & $711.226^{+74.913}_{-65.271}$   & $-4.768^{+0.235}_{-0.213}$  & & $13305.650^{+5542.198}_{-5336.604}$  & $-0.028^{+0.575}_{-0.788}$ & $0.168^{+0.640}_{-0.132}$ & $-26.058^{+3.587}_{-4.268}$   \\ \\
 B1 & & $1208.593^{+83.950}_{-73.195}$  & $-3.043^{+0.139}_{-0.124}$  & & $19478.925^{+4357.710}_{-5010.645}$  & $0.991^{+0.283}_{-0.396}$  & $0.073^{+0.470}_{-0.063}$ & $-27.226^{+4.746}_{-4.702}$   \\ \\
 B2 & & $496.428^{+56.103}_{-55.199}$   & $-5.067^{+0.211}_{-0.224}$  & & $12666.811^{+6209.063}_{-5140.843}$  & $-0.353^{+0.562}_{-0.729}$ & $0.260^{+0.703}_{-0.205}$ & $-21.666^{+2.930}_{-3.429}$   \\ \\
 C1 & & $2346.262^{+85.068}_{-87.401}$  & $-1.293^{+0.067}_{-0.069}$  & & $28895.523^{+3655.039}_{-3325.084}$  & $1.878^{+0.125}_{-0.132}$  & $0.110^{+0.411}_{-0.079}$ & $-27.095^{+3.842}_{-3.311}$   \\ \\
 D1 & & $1641.177^{+39.212}_{-38.121}$  & $-0.395^{+0.029}_{-0.030}$  & & $2958.208^{+436.720}_{-540.998}$ 	   & $0.042^{+0.103}_{-0.157}$  & $0.474^{+1.691}_{-0.394}$ & $-18.874^{+3.586}_{-4.360}$   \\ \\
 D2 & & $1201.541^{+31.622}_{-32.289}$  & $-0.195^{+0.031}_{-0.032}$  & & $3588.304^{+628.561}_{-548.905}$ 	   & $0.477^{+0.097}_{-0.104}$  & $0.244^{+0.748}_{-0.212}$ & $-13.518^{+2.312}_{-3.540}$   \\ \\
 E1 & & $1319.358^{+37.578}_{-37.398}$  & $-0.200^{+0.035}_{-0.036}$  & & $2182.169^{+489.394}_{-373.471}$ 	   & $0.156^{+0.137}_{-0.143}$  & $0.149^{+0.589}_{-0.121}$ & $-17.654^{+3.222}_{-3.758}$   \\  \\ 
 E2 & & $1499.589^{+49.260}_{-52.840}$  & $-0.056^{+0.056}_{-0.060}$  & & $2703.306^{+1529.192}_{-673.082}$    & $0.397^{+0.283}_{-0.315}$  & $0.321^{+1.636}_{-0.272}$ & $-18.332^{+4.736}_{-4.658}$   \\ \\
 \hline
& &  \multicolumn{7}{c}{ Set 2 } \\
\cline{3-9} \\
 A1 & & $465.315^{+67.145}_{-64.003}$  & $-6.378^{+0.309}_{-0.341}$  & & $11549.540^{+6273.411}_{-4609.940}$  & $-0.899^{+0.713}_{-0.912}$ & $0.219^{+0.562}_{-0.173}$ & $-25.601^{+3.310}_{-4.005}$   \\ \\
 B1 & & $860.176^{+77.479}_{-72.800}$  & $-4.489^{+0.192}_{-0.196}$  & & $15582.196^{+6354.234}_{-5239.984}$  & $0.222^{+0.528}_{-0.640}$  & $0.082^{+0.419}_{-0.067}$ & $-28.497^{+4.631}_{-4.533}$   \\ \\
 B2 & & $292.200^{+54.307}_{-49.132}$  & $-7.177^{+0.362}_{-0.374}$  & & $10678.567^{+5518.909}_{-5017.907}$  & $-1.409^{+0.733}_{-1.067}$ & $0.185^{+0.439}_{-0.154}$ & $-23.931^{+2.776}_{-4.149}$   \\ \\
 C1 & & $1782.191^{+90.391}_{-87.494}$ & $-2.443^{+0.104}_{-0.106}$  & & $42948.612^{+1406.500}_{-2885.071}$  & $2.324^{+6.049}_{-0.082}$  & $0.123^{+1.662}_{-0.103}$ & $-28.266^{+14.958}_{-4.754}$  \\ \\
 D1 & & $1781.551^{+53.061}_{-52.769}$ & $-0.984^{+0.049}_{-0.046}$  & & $9876.733^{+2103.361}_{-2116.730}$   & $0.703^{+0.177}_{-0.227}$  & $1.165^{+1.689}_{-0.870}$ & $-17.172^{+2.167}_{-3.118}$   \\ \\
 D2 & & $1500.575^{+43.331}_{-43.481}$ & $-0.658^{+0.036}_{-0.039}$  & & $6266.866^{+802.237}_{-960.549}$ 	   & $0.490^{+0.099}_{-0.126}$  & $0.321^{+1.201}_{-0.230}$ & $-16.770^{+2.846}_{-2.602}$   \\ \\
 E1 & & $1562.164^{+55.339}_{-56.103}$ & $-0.501^{+0.056}_{-0.059}$  & & $4048.862^{+1288.381}_{-887.903}$    & $0.395^{+0.237}_{-0.249}$  & $0.296^{+1.366}_{-0.234}$ & $-19.595^{+3.917}_{-3.846}$   \\ \\
 E2 & & $1828.049^{+38.017}_{-40.772}$ & $-0.390^{+0.027}_{-0.028}$  & & $2878.578^{+334.665}_{-273.075}$ 	   & $-0.045^{+0.081}_{-0.085}$ & $0.213^{+0.695}_{-0.177}$ & $-23.519^{+3.927}_{-4.950}$   \\ \\
 \hline
& &  \multicolumn{7}{c}{ Set 3 } \\
\cline{3-9} \\
 A1 & & $516.755^{+26.555}_{-12.364}$  & $-6.514^{+0.116}_{-0.095}$  & & $10300.418^{+6770.433}_{-5465.892}$  & $-1.658^{+0.865}_{-1.305}$ & $0.202^{+0.475}_{-0.167}$ & $-26.107^{+3.432}_{-4.605}$   \\ \\
 B1 & & $515.017^{+72.706}_{-62.625}$  & $-6.200^{+0.292}_{-0.307}$  & & $12761.177^{+6538.767}_{-5271.647}$  & $-0.700^{+0.665}_{-0.940}$ & $0.171^{+0.582}_{-0.139}$ & $-26.735^{+4.002}_{-4.106}$   \\ \\
 B2 & & $232.020^{+52.777}_{-45.074}$  & $-8.203^{+0.441}_{-0.495}$  & & $10327.801^{+6217.503}_{-5447.689}$  & $-2.007^{+0.883}_{-1.317}$ & $0.232^{+0.595}_{-0.192}$ & $-24.138^{+2.993}_{-4.419}$   \\ \\
 C1 & & $1331.405^{+83.233}_{-79.425}$ & $-3.505^{+0.137}_{-0.144}$  & & $52100.924^{+2082.091}_{-4800.979}$  & $2.404^{+0.093}_{-0.153}$  & $0.200^{+0.529}_{-0.141}$ & $-28.804^{+3.760}_{-3.259}$   \\ \\
 D1 & & $1427.909^{+62.389}_{-56.541}$ & $-1.959^{+0.072}_{-0.074}$  & & $11813.293^{+4826.088}_{-3755.064}$  & $0.547^{+0.374}_{-0.421}$  & $0.900^{+1.334}_{-0.734}$ & $-18.801^{+2.309}_{-4.469}$   \\ \\
 D2 & & $1475.761^{+53.870}_{-55.292}$ & $-1.243^{+0.055}_{-0.057}$  & & $11909.341^{+2525.662}_{-3102.992}$  & $0.822^{+0.177}_{-0.277}$  & $0.441^{+1.291}_{-0.356}$ & $-17.475^{+2.422}_{-3.616}$   \\ \\
 E1 & & $1889.145^{+62.948}_{-63.660}$ & $-0.605^{+0.057}_{-0.058}$  & & $18594.745^{+2743.095}_{-6819.408}$  & $1.728^{+0.111}_{-0.422}$  & $0.709^{+1.453}_{-0.566}$ & $-17.832^{+2.475}_{-3.077}$   \\ \\
 E2 & & $1698.964^{+47.724}_{-53.039}$ & $-1.007^{+0.043}_{-0.044}$  & & $3154.249^{+518.230}_{-423.661}$     & $-0.418^{+0.145}_{-0.149}$ & $0.203^{+0.724}_{-0.163}$ & $-24.623^{+0.163}_{-4.584}$   \\
 \hline
\end{tabular}
\caption{\label{tab:params_results} Best-fit parameters for the one-term and two-terms models of mass-loss curves for our simulated GCs in set 1, set 2 and set 3.}
\end{center}
\end{table*}

\begin{table*}
\centering
\begin{center}
\begin{tabular}{lcccc}
 \hline
   & \multicolumn{2}{c}{ AIC }  & \multicolumn{2}{c}{ BIC } \\
GC &  One-Term Model & Two-Terms Model & One-Term Model & Two-Terms Model \\
\hline
\hline
& \multicolumn{4}{c}{Set 1}\\
\cline{2-5}
A1 & 28.797  & 13.805 & 44.735  & 45.680 \\
B1 & 102.797 & 16.446 & 119.004 & 48.860 \\
B2 & 23.453  & 11.572 & 39.661  & 43.987 \\
C1 & 416.431 & 30.590 & 432.787 & 63.301 \\
D1 & 40.244  & 15.262 & 56.552  & 47.878 \\
D2 & 58.734  & 15.438 & 74.403  & 46.778 \\
E1 & 26.469  & 17.798 & 41.722  & 48.304 \\
E2 & 16.240  & 19.115 & 30.552  & 47.738 \\
\hline
& \multicolumn{4}{c}{Set 2}\\
\cline{2-5}
A1 & 14.715  & 11.640 & 30.795  & 43.800 \\
B1 & 49.655  & 13.927 & 65.868  & 46.352 \\
B2 & 10.189  & 10.479 & 26.402  & 42.904 \\
C1 & 340.34  & 36.038 & 356.700 & 68.760 \\
D1 & 115.281 & 19.164 & 131.594 & 51.791 \\
D2 & 108.604 & 14.629 & 124.869 & 47.160 \\
E1 & 34.736  & 17.341 & 49.884  & 47.637 \\
E2 & 48.049  & 20.349 & 64.522  & 53.297 \\
\hline
& \multicolumn{4}{c}{Set 3}\\
\cline{2-5}
A1 & 11.823  & 11.562 & 27.902  & 43.721 \\
B1 & 23.844  & 12.748 & 40.055  & 45.171 \\
B2 & 6.286   & 10.407 & 22.498  & 42.831 \\
C1 & 214.044 & 43.588 & 230.404 & 76.309 \\
D1 & 67.307  & 16.392 & 83.618  & 49.014 \\
D2 & 119.347 & 13.438 & 135.612 & 45.967 \\
E1 & 234.017 & 35.887 & 249.581 & 67.015 \\
E2 & 40.872  & 17.919 & 57.345  & 50.865 \\
\hline
\end{tabular}
\caption{\label{tab:model_comparison} Comparison of AIC and BIC values for one and two-term mass-loss models in our simulated GCs across sets 1, 2, and 3.}
\end{center}
\end{table*}


\bsp	
\label{lastpage}
\end{document}